\documentclass[journal,twocolumn]{IEEEtran}
\usepackage{cite}
\usepackage{amsmath,amsfonts,amsxtra,amssymb,latexsym,amscd,amsthm,mathrsfs,bm}
\usepackage{graphicx}
\graphicspath{{./figures/}}
\usepackage{hyperref}%%% this is giving me warning
\usepackage{xcolor}
\usepackage{algorithm}
\usepackage{algpseudocode}

\usepackage[small]{caption} 
\usepackage{epstopdf}% convert eps figure automatically to pdf for pdflatex
\usepackage{comment}
\usepackage{mathtools}
\newcommand*{\Scale}[2][4]{\scalebox{#1}{$#2$}}%

\newcommand{\RomanNumeralCaps}[1]
    {\MakeUppercase{\romannumeral #1}}

\makeatletter
\def\BState{\State\hskip-\ALG@thistlm}

\begin{document}

\title{Joint Superimposed Pilot-aided Channel Estimation and Data Detection for FTN Signaling over Doubly-Selective Channels}

\author{\IEEEauthorblockN{Simin Keykhosravi and Ebrahim Bedeer}\\
\IEEEauthorblockA{{Department of Electrical and Computer Engineering, University of Saskatchewan, Saskatoon, Canada} \\
Emails:{\{sik904, e.bedeer\}}@usask.ca}}

\maketitle
\begin{abstract} 

Faster-than-Nyquist (FTN) signaling and superimposed pilot (SP) techniques are effective solutions for significantly enhancing the spectral efficiency (SE) in next-generation wireless communication systems.
This paper proposes an innovative SP-aided channel estimation method for FTN signaling enhancing the SE {over} doubly-selective (i.e., time- and frequency-selective) channels.  
To avoid complex channel tracking, we utilize a basis expansion model (BEM) {to characterize doubly-selective channel variations.} 
We propose a frame structure that superimposes a known periodic pilot sequence {onto} the information sequence, avoiding SE loss by {eliminating the additional overhead of} multiplexed pilot (MP). 
Additionally, we find the optimal FTN signaling SP sequence that minimizes the mean square error (MSE) of {the channel estimation}.
Expanding on our proposed {SP-aided channel estimation} method, we propose {two detection methods: (1) an SP-aided separate channel estimation and data detection (SCEDD) method performing a single channel estimation followed by iterative data detection via a turbo equalizer, serving as a baseline for evaluating the SP-aided channel estimation method, and (2) an SP-aided joint channel estimation and data detection (JCEDD) method, which extends the SCEDD by updating the channel estimate in each turbo equalization iteration, becoming our primary focus for its superior performance.}
{At equivalent SE and a higher fading rate on the order of $10^{-3}$, our simulations show that SP-aided SCEDD method outperforms MP-aided techniques in both MSE and BER, while the SP-aided JCEDD method delivers remarkable performance, where reference approaches fail to track rapid channel variations.}
{At a very low fading rate on the order of $10^{-4}$, the SP-aided JCEDD algorithm enhances MSE by over 6 dB and 2 dB compared to the MP-aided methods in Ishihara and Sugiura (2017) and Keykhosravi and Bedeer (2023), respectively. In terms of BER, the JCEDD provides over 3 dB enhancements compared to Ishihara and Sugiura (2017), while remaining competitive with Keykhosravi and Bedeer (2023), showing only less than 0.5 dB degradation.}

\end{abstract}


\section{Introduction}
With the rapid increase in data traffic and limited bandwidth becoming a pressing concern in wireless communication systems, enhancing the spectral efficiency (SE) has become crucial. Faster-than-Nyquist (FTN) signaling, first proposed in the 1970s, has recently captured researchers' interest due to its potential to boost the SE without requiring additional bandwidth or increased energy per bit \cite{anderson2013faster}.

{The traditional Nyquist transmission, which employs $T$-orthogonal pulses over a symbol duration $T$, is constrained by the maximum rate ($1/T$). This limit guarantees inter-symbol interference (ISI)-free communication over frequency flat channels \cite{john2008digital}. On the contrary, FTN signaling employs $T$-orthogonal pulses to transmit data symbols at symbol intervals $\tau T$, which are shorter than the Nyquist limit. Here, $0 < \tau \leq 1$ denotes the FTN signaling acceleration parameter. While surpassing the Nyquist limit results in unavoidable ISI at the receiver, it significantly improves the SE.
Mazo demonstrated that the minimum squared Euclidean distance (MED) for uncoded binary sinc pulse transmission remains unchanged as $\tau$ is reduced to 0.802 \cite{mazo1975faster}, %This threshold, later termed the Mazo limit, indicates 
indicating that FTN signaling can achieve a $25\%$ times higher rate compared to conventional Nyquist signaling while maintaining the same energy per bit and bandwidth. If $\tau$ falls
below the Mazo limit, the MED decreases, and the improved SE comes at the cost of increased ISI, requiring effective ISI mitigation techniques at the receiver to maintain reliable detection \cite{anderson2013faster}.}

Most of the existing research on the detection of FTN signaling has focused on the transmission of the FTN signal in the presence of additive white Gaussian noise (AWGN), e.g.,\cite{liveris2003exploiting, prlja2008receivers, bedeer2017reduced,sugiura2013frequency,bedeer2017very,li2020beyond,ibrahim2021novel}.
The uncoded transmission of FTN signaling introduces ISI, which exhibits a trellis structure as demonstrated by \cite{liveris2003exploiting}. Thus, a trellis decoder can be used for data detection to address the effects of this trellis-structured ISI.
Because the optimal FTN signaling detection for small values of 
$\tau$ involves high computational complexity, even when AWGN is present, various studies have suggested using reduced trellis or reduced tree search methods as approximations to the optimal solution\cite{liveris2003exploiting, prlja2008receivers, bedeer2017reduced}. Additionally, tree search and trellis-based equalizers pose challenges when applied in scenarios with high modulation orders, primarily due to their inherently high computational complexity.
In scenarios involving high-order and ultra-high-order modulations, where FTN signaling detection becomes particularly challenging, effective strategies are introduced to reduce the computational
complexity, e.g., precoding techniques \cite{li2020beyond} and optimization methods utilizing the alternating directions multiplier method (ADMM) \cite{ibrahim2021novel}.

FTN signaling over AWGN channels in its early studies employs time-domain optimal detectors to mitigate ISI \cite{prlja2008receivers}. In scenarios where 
$\tau$ values are high, methods such as frequency domain equalization (FDE) \cite{sugiura2013frequency,sugiura2014frequency,li2020time} %,hong2016performanceor 
symbol-by-symbol detection \cite{bedeer2017very} are utilized to develop low-complexity detectors for FTN signaling. The original proposition of the frequency-domain FTN receiver was credited to Sugiura \cite{sugiura2013frequency}. In the work in \cite{sugiura2013frequency}, inserting a cyclic prefix in each transmitted block and approximating the FTN-induced ISI by a finite-tap circulant matrix structure paves the way to utilize an efficient fast Fourier transform operation along with a low-complexity channel-inverse-based minimum mean square error (MMSE) detection algorithm. The time and frequency domain equalizers for FTN signaling are compared in \cite{li2020time}, demonstrating that FDE is ineffective in FTN signaling detection when the packing factor is low. In contrast, time domain equalization remains reliable under these conditions.
The work in \cite{bedeer2017very} proposes an innovative successive symbol-by-symbol sequence estimator, estimating the transmitted symbols in a low-complexity manner.

In wireless communications, channel estimation can be accomplished through the utilization of pilot symbols known to the receiver. The primary concept behind the conventional multiplexed pilot (MP) transmission involves the time-division multiplexing of pilot symbols with the data. A few papers have explored techniques for estimating the channel impulse response in FTN signaling over frequency-selective channels \cite{hirano2014tdm, wu2017hybrid, ishihara2017iterative, shi2017frequency}. The work in \cite{hirano2014tdm} employed an MP approach, where the pilot is designed using the Nyquist criteria and placed before the data blocks to estimate the channel impulse response. In contrast to \cite{hirano2014tdm} that multiplexes Nyquist-based pilot sequence with data, the work in \cite{wu2017hybrid} utilized an MP sequence based on the FTN signaling criterion to perform time-domain joint channel estimation and FTN signaling detection. To address the complexity challenge in FTN signaling detection, the authors in \cite{ishihara2017iterative} proposed a low complexity joint channel estimation and data detection for FTN signaling in the frequency domain. However, this method introduces overhead to mitigate ISI contamination of the FTN signaling pilot sequence by inserting a guard block of the same length as the pilot sequence before the transmitted data within the frame. The work in \cite{ishihara2017iterative} also employs a cyclic prefix at the end of the transmitted data within the frame. This cyclic prefix has the same length as the pilot sequence and the guard block. Introducing an overhead of four times the pilot sequence length ultimately reduces the SE in \cite{ishihara2017iterative}. 

Most of the research on detecting FTN signaling over frequency-selective channels assumes the channel to be quasi-static \cite{hirano2014tdm, wu2017hybrid, ma2021generalized,ishihara2017iterative}. %. 
Conversely, the work in \cite{shi2017frequency} addresses a doubly-selective (i.e., time- and frequency-selective) channel. In \cite{shi2017frequency}, a method for frequency-domain joint channel estimation and data detection for FTN signaling is introduced, utilizing the message-passing algorithm. Recently, in \cite{keykhosravi2023pilot}, a novel method for channel estimation in FTN signaling over doubly-selective channels is {introduced.} 
{Unlike} the approach in \cite{ishihara2017iterative}, the frame structure proposed in \cite{keykhosravi2023pilot} achieves a better SE by eliminating the need for a cyclic prefix, which reduces overhead. It is worth mentioning that, on contrary to the methods focusing on the receiver of the FTN signaling system, various studies have focused on the utilization of the transmission techniques centered around precoding, as explored in \cite{li2020joint,wen2022joint,kang2020precoding}.

As mentioned earlier, the conventional MP method has been employed for channel estimation by multiplexing the pilot symbols with the data. However, the MP approach leads
to an undesirable reduction in data rate and compromises the SE in both Nyquist and FTN signaling studies. Hence, it is essential to adopt methods to further improve the SE.
Drawing upon insights from the principles of index modulation (IM), a recent paper \cite{keykhosravi2024} proposes an innovative IM-based {high frequency (HF)} channel estimation for FTN signaling to enhance the SE { by utilizing pilot sequence locations to transmit additional information bits.} 
{Existing MP‐aided methods in FTN signaling over doubly-selective channels \cite{keykhosravi2023pilot, keykhosravi2024, wu2016frequency} are designed for scenarios with low fading rates on the order of $10^{-4}$, where the fading rate is defined as the Doppler frequency shift normalized by the symbol rate. These methods require pilot sequences longer than the effective ISI from both the channel and FTN signaling, making them impractical as the fading rate increases, requiring more frequent pilot insertion for reliable channel estimation.}

In addition to the MP-aided method for wireless communication channel estimation, the superimposed pilot (SP) method was introduced as early as 1987 \cite{makrakis1987novel} for Nyquist signaling and {has been widely studied \cite{farhang1996experimental, hoeher1999channel,zhou2001superimposed,zhou2003first,tugnait2003channel,orozco2004channel,ghogho2004improved,tugnait2007doubly, he2008doubly}. The SP-aided methods provide an alternative approach, potentially mitigating rate loss by superimposing known pilots onto data, allowing full utilization of available time resources for data transmission.} 
{Early SP-based approaches \cite{zhou2001superimposed,zhou2003first,tugnait2003channel,orozco2004channel,ghogho2004improved} rely on superimposing a periodic training sequence, leveraging cyclostationary characteristics to estimate the channel using first- or second-order statistics of the received signal. These methods assume a time-invariant or block-fading channel. However, in rapidly time-varying conditions, the cyclostationary assumption may no longer hold, reducing estimation accuracy. 
{To address data and pilot interference, the authors in \cite{ghogho2005channel}, introduced a data-dependent superimposed pilot (DDSP) scheme where the training sequence consists of a known pilot and a data-dependent component unknown to the receiver. This design improves channel estimation accuracy by mitigating interference from unknown data but introduces data distortion. To address this issue%and enhance BER performance
, various techniques have been proposed, including interference cancellation \cite{ghogho2009full}, data identifiability improvements \cite{whitworth2007data}, and precoding-based elimination of data distortion \cite{chan2015elimination}. However, the studies in \cite{ghogho2005channel,ghogho2009full,whitworth2007data,chan2015elimination} assume the channel to be time-invariant or block-fading.}
To extend SP-based methods to doubly-selective channels, studies such as \cite{ tugnait2007doubly, he2008doubly} have employed basis expansion model (BEM) to track time-varying channels. 
SP-aided methods are particularly advantageous as the fading rate increases, where more frequent pilot insertion would otherwise be required for accurate channel estimation. 
In FTN signaling, where ISI is intentionally introduced to improve SE, conventional MP-aided methods require a pilot sequence longer than the effective ISI of both the channel and FTN signaling. This makes MP inefficient as the fading rate increases, necessitating more frequent pilot insertion. Thus, SP-aided methods are promising for enhancing the SE in FTN signaling, especially in such conditions. However, SP comes at the cost of introducing interference between the pilot and data, posing a potential challenge to channel estimation and data detection in FTN signaling.}

Inspired by SP techniques, we propose a novel SP-aided channel estimation method for FTN signaling, aimed at further enhancing the SE by removing the time slots allocated to pilot sequences. 
Expanding upon our proposed SP-aided doubly-selective channel estimation method for FTN signaling, we present an SP-aided separate channel estimation and data detection (SCEDD) method.
Furthermore, we propose an SP-aided joint channel estimation and data detection (JCEDD) method for FTN signaling to further improve both bit error rate (BER) and MSE.
The main contributions of this paper can be outlined as follows:

\begin{itemize}
\item 
We first propose an SP-aided doubly-selective channel estimation approach for FTN signaling by superimposing {a carefully chosen periodic pilot sequence onto the data sequence, facilitating channel estimation as detailed in the subsequent sections.} We adopt a frame structure that avoids SE loss by eliminating the need for an additional overhead of multiplexed pilots. { This becomes particularly critical as the fading rate increases (defined
as the Doppler frequency shift normalized by the symbol rate), requiring more MPs to maintain acceptable channel estimation accuracy. Additionally, the combined ISI from both the channel and FTN
signaling necessitates a longer MP to ensure reliable detection, which increases the overhead and reduces the SE.}
To avoid the complexity of channel tracking, we employ a BEM addressing the time-varying nature of doubly-selective channels. This is achieved by expressing the time-varying tap weights as a linear superposition of basis functions with time-invariant coefficients. Our proposed SP-aided channel estimation method for FTN signaling estimates the unknown BEM coefficients. 
More specifically {it} first computes the normalized discrete-time Fourier transform (DTFT) of the received signal, {as the magnitude at expected frequencies is anticipated} to be a function of the unknown BEM coefficients. Having the magnitude at the {expected} frequencies in normalized DTFT of the received signal, we employ the LSSE method to estimate the BEM coefficients in the time domain at the receiver. Finally, we estimate the doubly-selective channel coefficients having the estimated BEM coefficients.
We also find the optimal FTN signaling SP sequence that minimizes the MSE of doubly-selective channel estimation.

\item 

Building on our proposed SP-aided doubly-selective channel estimation method for FTN signaling, we introduce an SP-aided {detection framework with two approaches: SP-aided SCEDD and JCEDD. The SP-aided SCEDD provides a benchmark for evaluating the performance of our proposed SP-aided channel estimation method, where the channel is estimated once, followed by iterative data detection using a turbo equalizer. However, its performance remains limited due to the lack of iterative channel updates. 
To overcome these limitations, SP-aided JCEDD extends SP-aided SCEDD by refining the channel estimate in each iteration of the turbo equalizer, significantly enhancing data detection quality.}
Particularly, in each iteration of the turbo equalizer, the newly updated channel {estimate, derived from the detected data bits in the previous iteration, enhances the accuracy of decoding in the subsequent iteration.}
{Unlike approaches that complete the full data detection process (e.g., after multiple iterations of equalization) before updating the channel, JCEDD updates the channel estimate within each iteration of the turbo equalizer, aiming for a balanced trade-off between performance and computational complexity.}
{Extensive MSE and BER} simulation results demonstrate that, at comparable SE, the proposed SP-aided SCEDD approach outperforms the MP-aided {methods presented in \cite{keykhosravi2023pilot} and \cite{ishihara2017iterative}} for higher fading rates in the order of $10^{-3}$. 
{However, JCEDD, which builds on SCEDD by incorporating iterative channel estimation within the turbo equalizer, further enhances performance and outperforms not only the SP-aided SCEDD but also surpasses the MP-aided methods in \cite{keykhosravi2023pilot} and \cite{ishihara2017iterative}, where these techniques are unable to track higher channel variations associated with higher fading rates. 
At a lower fading rate in the order of $10^{-4}$, our SP-aided JCEDD algorithm improves MSE by more than 6 dB and 2 dB compared to \cite{keykhosravi2023pilot} and \cite{ishihara2017iterative}, respectively. Its BER performance shows a 3 dB gain over \cite{ishihara2017iterative} while remaining} on par with the MP-aided FTN system in \cite{keykhosravi2023pilot}, exhibiting less than 0.5 dB degradation.

\end{itemize}

The remainder of this paper is organized as follows. Section~\RomanNumeralCaps{2} introduces the system model of the proposed SP-aided channel estimation for FTN signaling. Our SP-aided method to estimate doubly-selective channels for FTN signaling is proposed in Section ~\RomanNumeralCaps{3}.  
Section ~\RomanNumeralCaps{4} presents our SP-aided SCEDD method for FTN signaling.
Moreover, Section ~\RomanNumeralCaps{4} details our proposed SP-aided JCEDD approach for FTN signaling. The simulation results are detailed in Section~\RomanNumeralCaps{5}. The final section, Section~\RomanNumeralCaps{6}, provides brief conclusions drawn from the previous sections of the paper.

Throughout the paper, matrices are denoted by boldface uppercase letters (e.g., $\mathbf{X}$), column vectors by boldface lowercase letters (e.g., $\mathbf{x}$), and scalars by lightface lowercase letters (e.g., $x$).
Matrix $\mathbf{X}$ is denoted as $\mathbf{X}^{-1}$ for its inverse, $\mathbf{X}^\text{T}$ for its transpose, and $\mathbf{X}^\text{H}$ for its complex conjugate transpose (Hermitian transpose). The notations $\mathbf{x} \ast \mathbf{y}$ and $\mathbf{x}  \circledast \mathbf{y}$ represent the convolution and circular convolution operators between $\mathbf{x}$ and $\mathbf{y}$, respectively. The expectation operator of $\mathbf{x}$ is represented by $\text{E}(\mathbf{x})$. The trace operation of matrix $\mathbf{X}$ is denoted by $\text{tr}({\mathbf{X}})$. Moreover, $\mathbf{x}^f$ indicates normalized DTFT of time-domain vector $\mathbf{x}$.
The notation $\lceil{x}\rceil$ represents the ceil operator{, which} rounds the real value $x$ up to the nearest integer greater than or equal to $x$. {Similarly, the notation $\lfloor{x}\rfloor$ represents the floor operator, which rounds the real value $x$ down to the largest integer less than or equal to $x$.
The modulo operation of $a$ with respect to $b$ is defined as $a \bmod b = a - b\lfloor{a/b}\rfloor$.}
 Lastly, $\mathbb{C}^{N\times M}$ encompasses all complex matrices with dimensions $N\times M$.

\begin{figure*}[t]
\centering
\hspace{2cm} % Adjust the value as needed
\includegraphics[width=0.78\textwidth]{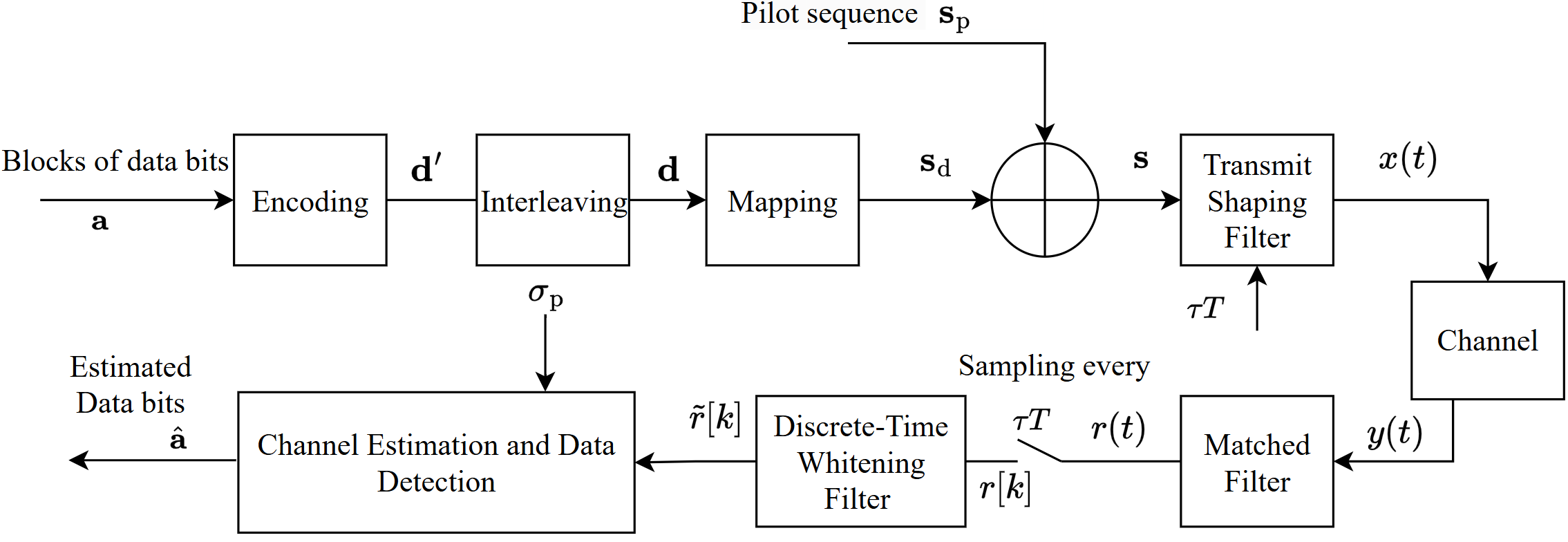}
\caption{The SP system model.}
\label{fig:1}
\end{figure*}
\begin{figure}[t]
\centering
\hspace{2cm} % Adjust the value as needed[width=0.85\textwidth] width=12cm,height=4.2cm
\includegraphics[width=7.2cm,height=3.5cm]{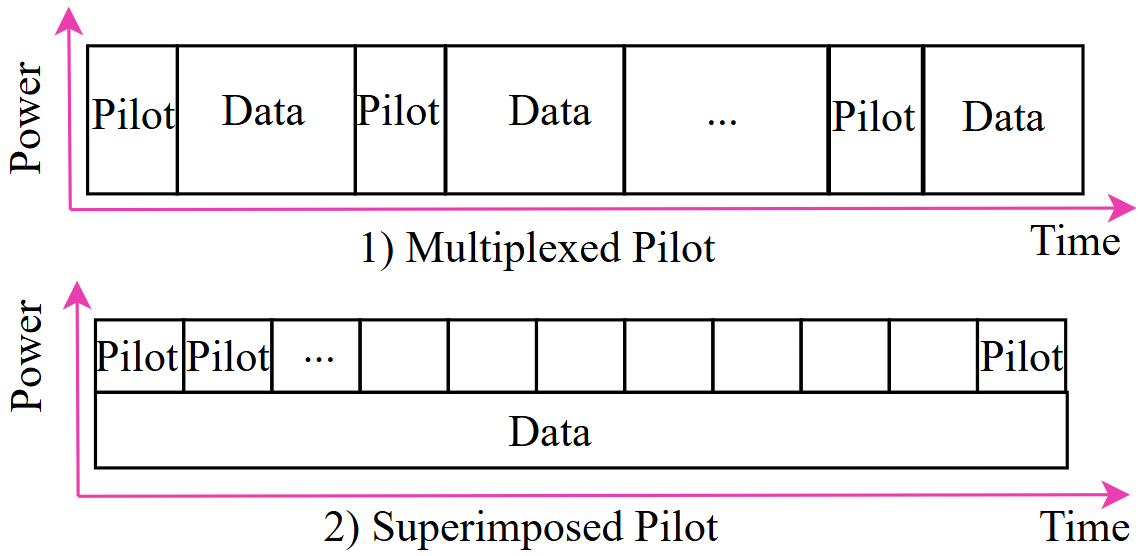} 
\caption{The SP-aided frame structure employed in the paper versus conventional MP-aided frame structure.}
\label{fig:11}
\end{figure}

\begin{figure*}[t]
\centering
\hspace{2cm} % Adjust the value as needed[width=0.85\textwidth] width=12cm,height=4.2cm
\includegraphics[width=11.4cm,height=4.7cm]{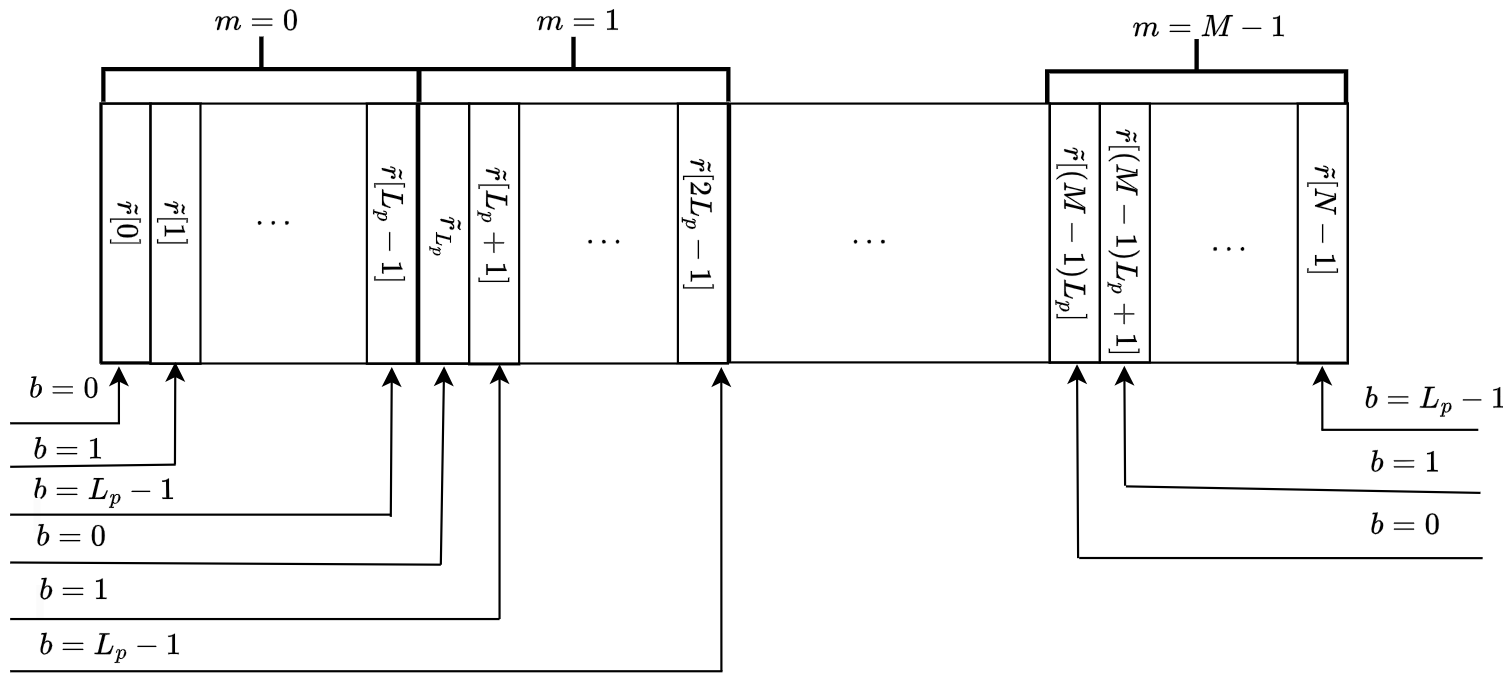}
\caption{The received signal structure employed for processing at the receiver.}
\label{fig:2}
\end{figure*}
\begin{figure}[t]
\centering
\hspace{2cm} % Adjust the value as needed
\includegraphics[width=7.3cm,height=3.3cm]{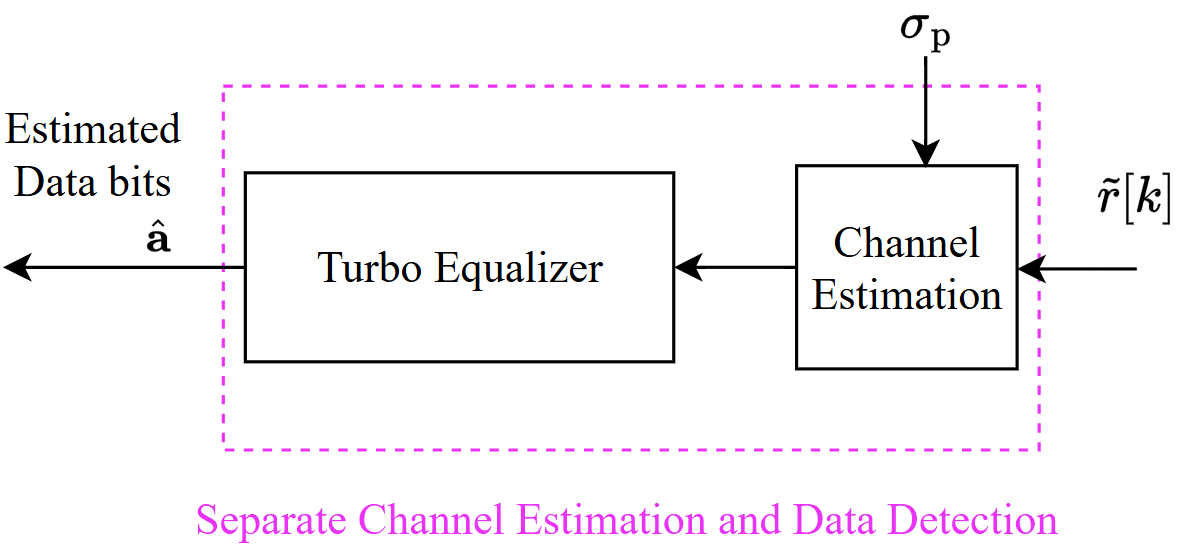}
\caption{The SP-aided SCEDD for the FTN signaling model.}
\label{fig:SCEDD}
\end{figure}
\begin{figure*}[t]
\centering
\hspace{2cm} % Adjust the value as needed
\includegraphics[width=0.5\textwidth]{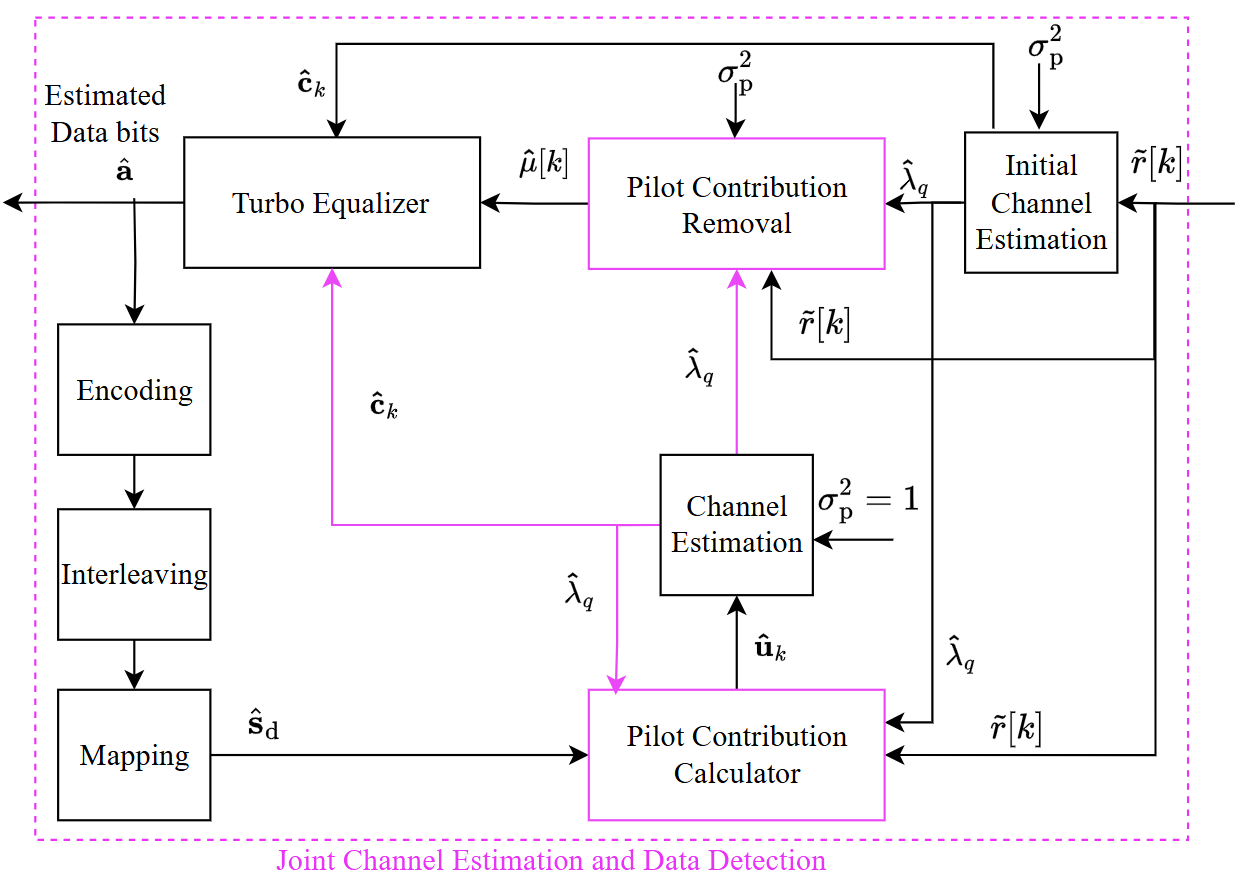}
\caption{The SP-aided JCEDD for the FTN signaling system model.}
\label{fig:JCEDD}
\end{figure*}

\section{System Model}

We consider the FTN signaling system based on superimposed pilots as depicted in Fig. \ref{fig:1}. At the transmitter end, every block of independent and identically distributed (i.i.d.) data bits denoted by $\textbf{a}$ undergoes encoding to yield $\textbf{d}'$. To counteract the impact of error bursts arising from fading channels, the interleaver rearranges the coded bit blocks into $\textbf{d}$. Subsequently, each set of ${\log_2}M$ consecutive output bits from the interleaver is mapped onto a complex data symbol, $s_{\text{d}}[n],\, n =0, 1, \ldots, N-1$, within a constellation set $\mathcal{S}_\text{d}$ for $M$-ary quadrature amplitude modulation ($M$-QAM), where $N$ is the frame length.

Facilitating channel estimation at the receiver, modulated symbols with known values, or pilots, $s_{\text{p}}[n],\, n =0, 1, \ldots, N-1$, are superimposed with data symbols to generate the transmitted symbols, $s[n] = \sigma_\text{d} s_{\text{d}}[n] + \sigma_\text{p} s_{\text{p}}[n]$, where $\sigma_\text{d}^2$ and $\sigma_\text{p}^2$ are the powers of data and pilot in the transmitted signal, respectively.
The SP-aided frame structure at the transmitter employed in this paper versus the conventional MP-aided frame structure is illustrated in Fig. \ref{fig:11}. 
%To guarantee that the received data symbols exhibit cyclostationary, aiding the channel estimation, 
We consider $s_{\text{p}}[n]$ as a periodic sequence with a period $L_\text{p}$ in a way that $N_\text{m} = \frac{N}{L_\text{p}}$ to be an integer, where $p_0[n],\, n =0, 1, \ldots, L_\text{p}-1$, is the pilot sequence in the first period of $s_{\text{p}}[n]$. 
After passing through a $T$-orthogonal root raised cosine (RRC) pulse denoted as $g(t)$ with unit energy, i.e., $\int_{-\infty}^{\infty} |g(t)|^2, dt = 1$, the symbols are transmitted at regular intervals of symbol duration $\tau T$. Having said that, the expression for the baseband transmitted signal can be formulated as 
\begin{IEEEeqnarray}{rcl}\label{equation1}
x(t) &{}={}& \sum\limits_{n=0}^{N-1} {s[n] g(t-n\tau T)}.
\end{IEEEeqnarray}
%%%%%%%%%%%%%%%%%%%%%%%%%%%%%%%%%%%%%%%%%%%%%%%%%%%%%%
For the transmission of FTN signaling over a doubly-selective fading channel, characterized as $c(t,\phi)$ the corresponding received signal is
\begin{IEEEeqnarray}{rcl}\label{equation2}
y(t) &{}={}& x(t)\ast c(t,\phi) + n(t).
\end{IEEEeqnarray}
Here, $n(t)$ represents the zero-mean AWGN having a double-sided power spectral density of $N_0/2$.
The received signal, $y(t)$, undergoes filtering with a filter matched to
$g(t)$, where $g^\ast(-t) = g(t)$. The signal at the output of the matched filter can be written as:
\begin{IEEEeqnarray}{rcl}\label{equation4}
r(t) &{}={}& x(t)\ast  c(t,\phi)\ast g(t)+z(t),
\end{IEEEeqnarray}
where $z(t) = n(t)\ast g(t)$. 
By replacing $s(t)$ from (\ref{equation1}), the expression for the received signal becomes
\begin{IEEEeqnarray}{rcl}\label{equation5}
r(t) =  \sum\limits_{n=0}^{N-1} {s[n] h(t-n\tau T)}\ast c(t,\phi)+z(t),
\end{IEEEeqnarray}
where $h(t) = g(t)\ast g(t)$ is the raised cosine (RC) pulse. While the theoretical ISI length of FTN signaling remains infinite, practical considerations arise from the observation that the power of the raised cosine pulse, $h(t)$, is primarily concentrated in and around its main lobe. As a result, we constrain the length of the raised cosine pulse, $h(t)$, to $2L_\text{h}+1$ so that the RC pulse has $2L_\text{h}+1$ taps. Thus we can represent the vector $\mathbf{h} =  \big[ h[-L_\text{h}],h[-L_\text{h}+1],\hdots,h[L_\text{h}] \big]^{\text{T}}$. The FTN signaling packing ratio $\tau$ and the roll-off factor of the RRC pulse $g(t)$ determine the parameter $L_\text{h}$. 

Utilizing the tapped delay line channel model \cite{cavers2006mobile}, we consider a doubly-selective channel characterized by the impulse response $c(t,\phi)$. This channel model incorporates $L_\text{c}$ channel taps, represented as $c_l(t)$ for $l = 0, 1, \ldots, {L_\text{c}-1}$, where each path $c_l(t)$ in the channel has a delay of $\phi_l (t)$. Sampling the matched-filtered signal $r(t)$ at intervals of $\tau T$, the $k$th received sample can be expressed as
\begin{IEEEeqnarray}{rcl}\label{equation6}
r[k] &{}={}& \sum\limits_{l=0}^{L_\text{c}-1}  c_l[k]  \sum\limits_{n=0}^{N-1} {s[n]}  h[k-n-l] +z[k],
\end{IEEEeqnarray}
where $c_l[k]=c_l(k\tau T)$. {Additionally, $h[k-n-l]$ represents the discrete-time sample of the truncated RC pulse $h(t)$, which is sampled at intervals \(\tau T\) and truncated to \(2L_h+1\) taps. The index $k-n-l$ accounts for the delays introduced by the $n$th symbol and the $l$th channel tap.}
To decorrelate the colored noise samples $z[k]$, the sampled signal in (\ref{equation6}) undergoes filtering by a discrete-time whitening filter, denoted as $\textbf{v}=\big[v[0], v[1], ..., v[L_\text{h}-1]\big]^\text{T}$ with a length of $L_\text{h}$, as explained in \cite{keykhosravi2023pilot}. The $k$th output sample of the whitening filter can be expressed as
\begin{IEEEeqnarray}{rcl}\label{equation71}
\tilde{r}[k] &{}={}& \sum\limits_{l=0}^{L_\text{c}-1}  c_l[k]  \sum\limits_{n=0}^{N-1} {s[n]}  v[k-n-l]  +w[k].
\end{IEEEeqnarray}
Here, $w[k]$ represents the zero-mean white Gaussian noise with power $\sigma_n^2$.

{In order to avoid the complexities associated with channel tracking, we employ the BEM to efficiently address the time-varying nature of channels by representing the time-varying tap weights as a linear superposition of basis functions having time-invariant coefficients \cite{rabbi2008basis}. Having said that, we write a discrete-time complex exponential basis expansion model for the doubly-selective fading channel as \cite{ma2003maximum}
\begin{IEEEeqnarray}{rcl}\label{equation9}
c_l[k]&{}={}&\sum\limits_{q=0}^{Q} \lambda_{q}[l] \exp({j \omega_q k}),
\end{IEEEeqnarray}
where $\omega_q = 2 \pi (q-Q/2)/N$, $Q=2 \lceil{f_dN\tau T} \rceil $ is the BEM order, and $\lambda_{q}[l]$ is the basis coefficient for path $l$ of the $q$th basis for $l = 0, ..., L_\text{c}-1$, ($\lambda_{q}[l]$ is zero for $l>L_\text{c}-1$). Here, $f_d$ is the maximum Doppler spread. It is worth mentioning that considering the BEM introduced in (\ref{equation9}), the problem of estimating the unknown doubly-selective channel coefficient $c_l[k]$, reduces to estimating $\omega_q$ and $\lambda_{q}[l]$ for path $l = 0, ..., L_\text{c}-1$ and $q = 0, ..., Q$. It is noteworthy that $\omega_q$ depends on the parameters  
$f_d$, $N$, $\tau$, and $T$ which are known at the receiver. Thus, $\omega_q$ can be considered known at the receiver.
Based on the BEM model in (\ref{equation9}), we rewrite (\ref{equation71}) as
\begin{IEEEeqnarray}{rcl}\label{equation72}
\small{\!\!\tilde{r}[k]}&{}={}&\!\!\! \small{\sum\limits_{l=0}^{L_\text{c}-1} \!\! \sum\limits_{q=0}^{Q} \! \lambda_{q}[l] \exp({j \omega_q k}) \!\! \sum\limits_{n=0}^{N-1}\!\! {s[n]}v[k-n-l]+w[k].}
\end{IEEEeqnarray} 
}

%%%%%%%%%%%%%%%%%%%%%%%%%%%%%%%%%%%%%%%%
\section{SP-aided Doubly-Selective Channel Estimation For FTN Signaling}
This section proposes an SP-aided method to estimate doubly-selective channels for FTN signaling. We also find the optimal {SP} sequence that minimizes the MSE of doubly-selective channel estimation. 
%%%%%%%%%%%%%%%%%%%%%%%%%%%%%%%%%%%%%%%%%%%%%%%%%%%%%%% 
As mentioned before $s[k]= \sigma_\text{d} s_{\text{d}}[k]+\sigma_\text{p} s_{\text{p}}[k]$. We define  $s_{\text{d}}[k]$ and $s_{\text{p}}[k]$ after passing through the whitening filter, respectively as $\tilde{s}_{\text{d}}[k]$ and $\tilde{s}_{\text{p}}[k]$ as
\begin{IEEEeqnarray}{rcl}\label{equation921}
\tilde{s}_{\text{d}}[k] &{}={}&    \sum\limits_{n=0}^{N-1} {s}_{\text{d}}[n] v[k-n], \nonumber\\
\tilde{s}_{\text{p}}[k] &{}={}&  \sum\limits_{n=0}^{N-1} {s}_{\text{p}}[n] v[k-n]. \nonumber\\
\end{IEEEeqnarray}
With (\ref{equation921}), the received signal in (\ref{equation72}) is reformulated as
\begin{IEEEeqnarray}{rcl}\label{equation73}
\tilde{r}[k] &{}={}& \sum\limits_{l=0}^{L_\text{c}-1} \sum\limits_{q=0}^{Q} \lambda_{q}[l] \exp({j \omega_q k})  (\sigma_\text{p}\tilde{s}_{\text{p}}[k-l]+\sigma_\text{d}\tilde{s}_{\text{d}}[k-l]) \nonumber\\
&{}+{}& w[k].
\end{IEEEeqnarray}
We define the received signal in (\ref{equation73}) to have two parts as 
\begin{IEEEeqnarray}{rcl}\label{equation91}
\tilde{r}[k]&{}=
    u[k]+\mu[k],
\end{IEEEeqnarray}
where 
\begin{IEEEeqnarray}{rcl}\label{equation92}
u[k] &{}={}&  \sigma_\text{p} \sum\limits_{q=0}^{Q} \exp({j \omega_q k}) \sum\limits_{l=0}^{L_\text{c}-1} \lambda_{q}[l]   \tilde{s}_{\text{p}}[k-l], 
\end{IEEEeqnarray}
\begin{IEEEeqnarray}{rcl}\label{equation93}
\mu[k] &{}={}& \sigma_\text{d} \sum\limits_{q=0}^{Q} \exp({j \omega_q k}) \sum\limits_{l=0}^{L_\text{c}-1} \lambda_{q}[l]   \tilde{s}_{\text{d}}[k-l] + w[k].
\end{IEEEeqnarray}
Thus, the problem of doubly-selective channel estimation from the received signal, $\tilde{r}[k]$, now becomes finding the unknown channel basis coefficients, $\lambda_q[l]$, $l = 0, ... L_\text{c}-1$, $q = 0, ..., Q$, having the known $\tilde{s}_{\text{p}}[k]$ and $\omega_q$ under the additive unknown term $\mu[k]$. 

As mentioned before, we consider $s_{\text{p}}[n]$ as a periodic sequence with a period $L_\text{p}$ to facilitate channel estimation as discussed later in detail. The length of the pilot should designed to be greater than the effective ISI length, i.e., $L_\text{p} \ge L_\text{c} + L_\text{h} -1$ to capture the ISI of both channel and FTN signaling.
%Now that we have $\tilde{s}_{\text{p}}[k]$ and ${s}_{\text{p}}[k]$ to be a periodic sequence with the same period $L_\text{p}$, 
We express $k=0, ..., N-1$ (in a block of length $N$) as $mL_\text{p} + b$ with $m=0,1, \hdots, N_\text{m}-1$ and a remainder $b=0,1, \hdots,L_\text{p}-1$. Hence, $\tilde{r}_b[m]$ denotes the $m$-th sub-block of length $L_\text{p}$ in the sequence $\tilde{r}[k]$, where $b$ is the index of the $b$th symbol within the sub-block. {The received signal structure employed for processing at the receiver is illustrated} in Fig. \ref{fig:2}. Applying (\ref{equation91}) allows us to decompose $\tilde{r}_b[m]$ into two parts as follows
\begin{IEEEeqnarray}{rcl}\label{equation917}
\tilde{r}_b[m]&{}= u_b[m]+\mu_b[m].
\end{IEEEeqnarray}
To determine $\tilde{r}_b[m]$ in (\ref{equation917}), we need to calculate $u_b[m]$ and $\mu_b[m]$.
We first calculate the $u_b[m]$ by writing $\tilde{s}_{\text{p}}[k]$ in (\ref{equation92}) according to (\ref{equation921}) as
\begin{IEEEeqnarray}{rcl}\label{equation102}
\tilde{s}_{\text{p}}[mL_\text{p} + b] &{}={}& \sum\limits_{r=0}^{L_\text{h}-1} {s}_{\text{p}}[mL_\text{p} + b-r] v[r]  \nonumber \\
&{}={}&  p_0[b] \circledast v[b],\nonumber \\
&{}\overset{\triangle}{=} &\tilde{p}_0[b],\IEEEeqnarraynumspace
\end{IEEEeqnarray}
where $\circledast$ is the circular convolution operator. The second equality arises from the periodicity of ${s}_{\text{p}}[k]$ with the period $L_\text{p}$.  
According to (\ref{equation102}), $\tilde{s}_{\text{p}}[k]$ is a periodic sequence with the same period $L_\text{p}$ as the sequence in ${s}_{\text{p}}[k]$. We define $\tilde{p}_0[b] = p_0[b] \circledast v[b],\, b =0, 1, \ldots, L_\text{p}-1$, to be the first period of $\tilde{s}_{\text{p}}[k]$. 

Substituting $\tilde{s}_{\text{p}}[k]$ from (\ref{equation102}) and $k = mL_\text{p}+b$ in (\ref{equation92}), the resulting expression unfolds as
\begin{IEEEeqnarray}{rcl}\label{equation11}
u_b[m]&{}={}& u[mL_\text{p} + b]\nonumber\\
&{}={}& \sum\limits_{q=0}^{Q} f_q[b] \exp({j L_\text{p} \omega_q  m}),
\end{IEEEeqnarray}
where 
\begin{IEEEeqnarray}{rcl}\label{equation13}
f_q[b]&{}={}& \sigma_\text{p} \exp({j \omega_q b}) \big(\lambda_q[b] \circledast \tilde{p}_0[b]\big).
\end{IEEEeqnarray}
Hence, for a given $b$, $u_b[m]$ in (\ref{equation11}) is composed of $Q+1$ complex exponentials in $m$, featuring complex amplitudes $f_q[b]$ and frequencies  $L_\text{p}\omega_q$.
%%%%%%%%%%%%%%%%%%%%%%%%%%%%%
%%%%%%%%%%%%%%%%%%%%%%%%%%%%%%%%%

Now that we have $u_b[m]$ from (\ref{equation11}), we need to find $\mu_b[m]$ in order to calculate $\tilde{r}_b[m]$ in (\ref{equation917}).
By substituting $k = mL_\text{p}+b$ into (\ref{equation93}), we obtain
\begin{IEEEeqnarray}{rcl}\label{equation112}
\mu_b[m]&{}={}& \mu[mL_\text{p} + b]\nonumber\\
&{}={}& \sigma_\text{d} \sum\limits_{q=0}^{Q} \exp({j \omega_q b}) \sum\limits_{l=0}^{L_\text{c}-1} \lambda_q[l] \exp({j L_\text{p} \omega_q  m}) \tilde{s}_{{\text{d}},{b-l}}[m]\nonumber\\
&{}+{}&w_b[m].
\end{IEEEeqnarray}

Considering the received signal $\tilde{r}_b[m]$ in (\ref{equation917}), the problem of doubly-selective channel estimation now becomes finding the unknown channel basis coefficients, $\lambda_q[l]$, $l = 0, ... L_\text{c}-1$, $q = 0, ..., Q$, having the known $\tilde{p}_0[b]$ and $\omega_q$ under the additive unknown term $\mu_b[m]$.

It should be noted that, on the one hand, from (\ref{equation11}),  
we infer that the magnitude of the normalized DTFT of ${u}_b[m]$ with respect to $m$ for a given $b$ is $f_q[b]$ around frequencies $\xi=L_\text{p}\omega_q$. 
As demonstrated in (\ref{equation13}), $f_q[b]$ is a function of the known $\omega_q$ and $\tilde{p}_0$ and unknown BEM coefficients, $\lambda_q[l]$, $l = 0, ... L_\text{c}-1$, $q = 0, ..., Q$. Thus having $f_q[b]$ makes it possible to estimate the unknown BEM coefficients, $\lambda_q$.
On the other hand, we expect the normalized DTFT of ${\mu}_b[m]$ in (\ref{equation112}) to be zero around $\xi=L_\text{p}\omega_q$ because the mean of the data and noise is zero as discussed later in this section. Having said that, we expect the magnitude of the normalized DTFT of received signal $\tilde{r}_b[m]$ (\ref{equation917}) around $\xi=L_\text{p}\omega_q$ to be $f_q[b]$. Finding the magnitude $f_q[b]$, makes it possible to estimate the unknown BEM coefficients, $\lambda_q$. Having an estimation of $\lambda_q[l]$, $l = 0, ... L_\text{c}-1$, $q = 0, ..., Q$, we can find the unknown channel coefficients, $c_l[k]$, $l = 0, ..., L_\text{c}-1$, from (\ref{equation9}). We provide a detailed discussion in the following subsections, where we introduce our proposed SP-aided method for doubly-selective channel estimation in FTN signaling, outlining the process step by step and identifying the optimal pilot sequence that minimizes the MSE in doubly-selective channel estimation.
%%%%%%%%%%%%%%%%%%%%%%%%%%%%%%
\subsection{Computing the normalized DTFT of the received signal}
First, we compute the normalized DTFT of the received signal, $\tilde{r}_b[m]$, with respect to $m$ for a given $b$ 
\begin{IEEEeqnarray}{rcl}\label{equation15}
\tilde{r}^f_b(\xi)&{}={}& \frac{1}{N_\text{m}} \sum\limits_{m=0}^{N_\text{m}-1} \tilde{r}_b[m] \exp({-j \xi m}).
\end{IEEEeqnarray}
Given (\ref{equation917}), we can express $\tilde{r}^f_b(\xi)$ as
\begin{IEEEeqnarray}{rcl}\label{equation151}
\tilde{r}^f_b(\xi)&{}={}& {u}^f_b(\xi)+{\mu}^f_b(\xi),
\end{IEEEeqnarray}
where ${u}^f_b(\xi)$ and ${\mu}^f_b(\xi)$ are the normalized DTFT of ${u}_b[m]$ ${\mu}_b[m]$ with respect to $m$ for a given $b$, respectively, and defined as
\begin{IEEEeqnarray}{rcl}\label{equation152}
{u}^f_b(\xi)&{}={}& \frac{1}{N_\text{m}} \sum\limits_{m=0}^{N_\text{m}-1} {u}_b[m] \exp({-j \xi m}), \nonumber\\
{\mu}^f_b(\xi)&{}={}& \frac{1}{N_\text{m}} \sum\limits_{m=0}^{N_\text{m}-1} {\mu}_b[m] \exp({-j \xi m}).
\end{IEEEeqnarray}
To find $\tilde{r}^f_b(\xi)$ in (\ref{equation151}), we need to calculate the ${u}^f_b(\xi)$ and ${\mu}^f_b(\xi)$. First, we find the ${u}^f_b(\xi)$. Having $u_b[m]$ from (\ref{equation11}), we can rewrite ${u}^f_b(\xi)$ in (\ref{equation152}) as 
\begin{IEEEeqnarray}{rcl}\label{equation153}
{u}^f_b(\xi)&{}={}& \sum\limits_{q=0}^{Q} f_q[b] \delta({\xi-L_\text{p} \omega_q }).
\end{IEEEeqnarray}
Since our focus is on the DTFT of the received signal around $\xi=L_\text{p}\omega_q$, from the (\ref{equation153}), we determine the value of ${u}^f_b(\xi)$ for a given $q$ at $\xi=L_\text{p}\omega_q$ as follows  
\begin{IEEEeqnarray}{rcl}\label{equation1531}
{u}^f_b(L_\text{p}\omega_q)&{}={}& f_q[b].
\end{IEEEeqnarray}
It should be noted that $f_q[b]$ is a function of known $\omega_q$ and $\tilde{p}_0[b]$ and unknown $\lambda_q[b]$ (channel basis coefficient) as demonstrated in (\ref{equation13}). We find the value of $f_q[b]$ around expected locations $\xi=L_\text{p}\omega_q$ and from which we can estimate the unknown $\lambda_q$. After estimating the channel basis coefficients, $\lambda_q[b]$, and having $\omega_q$, we can find the channel coefficients $c_l[k]$ from (\ref{equation9}).

With ${u}^f_b(\xi)$ determined from (\ref{equation153}), the next step is to calculate ${\mu}^f_b(\xi)$ to find $\tilde{r}^f_b(\xi)$ in (\ref{equation151}). Given $\mu_b[m]$ from (\ref{equation112}), we can express ${\mu}^f_b(\xi)$ in (\ref{equation152}) as
\begin{IEEEeqnarray}{rcl}\label{equation154}
{\mu}^f_b(\xi)&{}={}& \sigma_\text{d} \sum\limits_{q=0}^{Q} \exp({j \omega_q b}) \sum\limits_{l=0}^{L_\text{c}-1} \lambda_q[l] \tilde{s}^f_{{\text{d}},{b-l}}(\xi-L_\text{p} \omega_q ) + w^f_b(\xi)\nonumber\\
&{}={}& \!\! \sigma_\text{d} \!\!\sum\limits_{q=0}^{Q} \exp({j \omega_q b})\!\!  \left(\!\lambda_q[b] \ast \!\tilde{s}^f_{{\text{d}},{b-l}}(\xi-L_\text{p} \omega_q )\!\!\right)\!\!+ \! w^f_b(\xi).
\end{IEEEeqnarray} 
 %It should be noted 
According to the normalized DTFT definition in (\ref{equation15}), we can write $\tilde{s}^f_{{\text{d}},{b-l}}(\xi-L_\text{p} \omega_q )$ in (\ref{equation154}) at $\xi=L_\text{p}\omega_q$, $\tilde{s}^f_{{\text{d}},{b-l}}(0)$, as
\begin{IEEEeqnarray}{rcl}\label{equation161}
\tilde{s}^f_{{\text{d}},{b-l}}(0)&{}={}& \frac{1}{N_\text{m}} \sum\limits_{m=0}^{N_\text{m}-1} s_{\text{d},b-l}[m],
\end{IEEEeqnarray}
which is the mean of $s_{\text{d},b-l}[m]=s_{\text{d}}[k=mL_\text{p}+b-l]$. As the data is zero-mean, we conclude that $\tilde{s}^f_{{\text{d}},{b-l}}(\xi-L_\text{p} \omega_q )$ is zero at $\xi=L_\text{p}\omega_q$. That said, the value of ${\mu}^f_b(\xi)$ at $\xi=L_\text{p}\omega_q$ for a given $q$ in (\ref{equation154}) is
\begin{IEEEeqnarray}{rcl}\label{equation1541}
{\mu}^f_b(L_\text{p}\omega_q)&{}={}& w^f_b(L_\text{p}\omega_q).
\end{IEEEeqnarray} 
\subsection{ Employing the LSSE method to estimate the BEM coefficients}
As mentioned before, we are interested in the DTFT of the received
signal around $\xi=L_\text{p}\omega_q$. Now that we computed the $\tilde{r}^f_b(\xi)$ in (\ref{equation15}), we consider the $\tilde{r}^f_b(\xi)$ for a given $q$ at the location $\xi=L_\text{p}\omega_q$ having $b=0,1, \hdots,L_\text{p}-1$. %Substituting (\ref{equation153}) and (\ref{equation154}) in (\ref{equation151}) and setting $\xi=L_\text{p}\omega_q$, we have 
Substituting (\ref{equation1531}) and (\ref{equation1541}) in (\ref{equation151}) for $\xi=L_\text{p}\omega_q$, we have 
\begin{IEEEeqnarray}{rcl}\label{equation155}
\tilde{r}^f_b(L_\text{p} \omega_q )&{}\!=\!{}&  f_q[b] +  w^f_b(L_\text{p}\omega_q),\nonumber\\
&{}\!=\!{}& \sigma_\text{p} \exp({j \omega_q b}) \big(\lambda_q[b] \circledast \tilde{p}_0[b]\big)\! + \! w^f_b(L_\text{p}\omega_q).
\end{IEEEeqnarray}
The values of $\tilde{r}^f_b(L_\text{p} \omega_q )$ for, $b =0, 1, \ldots, L_\text{p}-1$, and a given $q$ in (\ref{equation155}) can be expressed in vector format as 
\begin{IEEEeqnarray}{rcl}\label{equation156}
\tilde{\mathbf{r}}^f_q&{}={}&  \sigma_\text{p} \mathbf{P}_{q}  \mathbf{V}  {\bm{\lambda}}_q +  \mathbf{w}^f_q,
\end{IEEEeqnarray}
where $\tilde{\mathbf{r}}^f_q=[\tilde{r}^f_0(L_\text{p} \omega_q ), \tilde{r}^f_1(L_\text{p} \omega_q ), \hdots \tilde{r}^f_{L_\text{p}-1}(L_\text{p} \omega_q )]^\text{T}$, $\textbf{w}^f_\text{q}=[w^f_0(L_\text{p}\omega_q), w^f_1(L_\text{p}\omega_q), \hdots, w^f_{L_\text{p}-1}(L_\text{p}\omega_q)]^\text{T}$, ${\bm{\lambda}}_q = \big[\lambda_q[0], \lambda_q[1], \hdots, \lambda_q[L_\text{c}-1]\big]^\text{T}$ and {the element in the $l$th-row and $k$th-column of matrix $\mathbf{P}_{q} \in \mathbb {C}^{L_\text{p}\times L_\text{p}}$ is given by ${p}_0[(l-k) \text{ mod } L_\text{p}] \exp({j \omega_q (l-1)})$.}
The matrix $\mathbf{V} \in \mathbb{C}^{L_\text{p} \times L_\text{c}}$ is the circulant FTN ISI matrix constructed from the elements of vector $\underset{\bar{}}{\textbf{v}}$ represented by $\underset{\bar{}}{\textbf{v}} =[v[0], v[1],..., v[L_\text{h}], 0,\hdots,0]^\text{T}$ of length $L_\text{p}$.

We employ the LSSE approach to estimate the unknown channel basis coefficients ${\bm{\lambda}}_q$, $q=0,1, ..., Q$. This entails minimizing the sum of squared errors as
\begin{IEEEeqnarray}{rcl}\label{equS5wh}
\bm{\hat{\lambda}}_q &{}={}& \text{arg} \displaystyle \min_{{\bm{\lambda}}_q}  (\tilde{\mathbf{r}}^f_q- \sigma_\text{p} \mathbf{P}_{q}  \mathbf{V}   {\bm{\lambda}}_q)^\text{H}(\tilde{\mathbf{r}}^f_q- \sigma_\text{p} \mathbf{P}_{q}  \mathbf{V}    {\bm{\lambda}}_q)\nonumber\\
 &{}={}& \frac{1}{ \sigma_\text{p}}(\mathbf{P}_{q}  \mathbf{V} )^{-1} \tilde{\mathbf{r}}^f_q.
\end{IEEEeqnarray}
The matrix $\mathbf{P}_{q}$ is determined by the known pilot sequence and $w_q$. The matrix $\mathbf{V}$ is also known at the receiver because the taps of the $\mathbf{v}$ are constant for an RRC pulse's roll-off factor and a given FTN signaling packing ratio. As a result, the matrix $(\mathbf{P}_{q}\mathbf{V})^{-1} $ can be calculated offline at the receiver. 

{After estimating} the channel basis coefficients, ${\bm{\hat{\lambda}}}_q$, in (\ref{equS5wh}), we can find $\mathbf{\hat{c}}_k = \big[\hat{c}_0[k], \hat{c}_1[k], \hdots, \hat{c}_{L_\text{c}-1}[k]\big]^\text{T}$ from (\ref{equation9}).
\subsection{Optimal pilot sequence design for SP-aided doubly-selective channel estimation}
Now that we have estimated the doubly-selective channel coefficients, $\mathbf{\hat{c}}_k$, we would like to find the optimal pilot that minimizes the MSE of SP-aided doubly-selective channel estimation. 
Relying on (\ref{equation9}), (\ref{equation156}), and (\ref{equS5wh}), we can write
\begin{IEEEeqnarray}{rcl}\label{equation160}
\mathbf{\hat{c}}_k-\mathbf{c}_k&{}={}& \sum\limits_{q=0}^{Q}  ({\bm{\hat{\lambda}}}_q-{\bm{\lambda}}_q ) \exp({j \omega_q k})\nonumber\\
 &{}={}& \frac{1}{ \sigma_\text{p}} \sum\limits_{q=0}^{Q} (\mathbf{P}_{q}\mathbf{V})^{-1} \mathbf{w}^f_q \exp({j \omega_q k}).
\end{IEEEeqnarray}
Given (\ref{equation160}), we can now calculate the MSE of channel estimation as
\begin{IEEEeqnarray}{rcl}\label{equS6wh1}
\text{MSE}
&{}={}& \text{E} \left( (\mathbf{\hat{c}}_k-\mathbf{c}_k)  (\mathbf{\hat{c}}_k-\mathbf{c}_k) ^\text{H}\right), \nonumber\\
 &{}={}& \Scale[0.95]{\frac{1}{ \sigma^2_\text{p}} \text{E}\!\left(\!  \sum\limits_{q=0}^{Q}\! \sum\limits_{{q'}=0}^{Q} \!({\bm{\hat{\lambda}}}_q-{\bm{\lambda}}_q )
({\bm{\hat{\lambda}}}_{q'}-\mathbf{\lambda}_{q'} )^\text{H} \exp({j (\omega_q-\omega_{q'})k}) \! \!\right)},\nonumber \\
 &{}={}& \frac{1}{ \sigma^2_\text{p}} \text{E}\left( \sum\limits_{q=0}^{Q} \sum\limits_{{q'}=0}^{Q} ((\mathbf{P}_{q}\mathbf{V})^{-1} \mathbf{w}^f_q )
((\mathbf{P}_{q}\mathbf{V})^{-1} \mathbf{w}^f_{q'} )^\text{H} ) \right)\nonumber \\
&{}\times{}&\exp({j (\omega_q-\omega_{q'})k}),\nonumber \\
&{}={}&  \frac{1}{ \sigma^2_\text{p}} \sum\limits_{q=0}^{Q} \sum\limits_{{q'}=0}^{Q} (\mathbf{P}_{q}\mathbf{V})^{-1}\text{E}\left( \mathbf{w}^f_q 
(\mathbf{w}^f_{q'} )^\text{H} \right) (\mathbf{P}_{q}\mathbf{V})^{-1} )^\text{H}\nonumber \\
&{}\times{}& \exp({j (\omega_q-\omega_{q'})k}) ,\nonumber \\
&{}={}&{\frac{\sigma_n^2}{ {L_\text{p}}.\sigma^2_\text{p}} \sum\limits_{q=0}^{Q}   \text{tr} \left( (\mathbf{V}^\text{H}\mathbf{P}_{q}^\text{H}\mathbf{P}_{q}\mathbf{V})^{-1}  \right)}. 
\end{IEEEeqnarray}
{With the definition of normalized DTFT of a vector in (\ref{equation15}), we have $\mathbf{w}^f_q =\frac{1}{\sqrt{{L_\text{p}}}} \mathbf{Q}_{L_\text{p}} \mathbf{w}_q$, where the matrix $\mathbf{Q}_{{L_\text{p}}} \in \mathbb{C}^{{L_\text{p}} \times {L_\text{p}}}$ is the normalized DTFT matrix for which the $i$th-row and $k$th-column is defined as $\frac{1}{\sqrt{{L_\text{p}}}}  \exp(\frac{-j2\pi i k}{{L_\text{p}}})$. Having said that $\text{E}\left( \mathbf{w}^f_q (\mathbf{w}^f_{q'} )^\text{H} \right) = \frac{1}{{L_\text{p}}} \mathbf{Q}_{L_\text{p}} \text{E} \left( \mathbf{w}_q (\mathbf{w}_{q'} )^\text{H} \right) \mathbf{Q}_{L_\text{p}}^\text{H}$ is zero for $q\neq q'$ and is $\frac{\sigma_n^2}{{L_\text{p}}} $ for $q= q'$.}
{As mentioned earlier, the element in the $l$th-row and $k$th-column of matrix $\mathbf{P}_{q}$ is given by ${p}_0[(l-k) \text{ mod } L_\text{p}] \exp({j \omega_q (l-1)})$. Thus, the element in $m$th-row and $k$th-column of matrix $\mathbf{P}_{q}^\text{H}\mathbf{P}_{q}$ can be written as}
{\begin{IEEEeqnarray}{rcl}\label{eqP21}
\mathbf{P}_{q}^\text{H}\mathbf{P}_{q} [m,k] &{}={}&  \sum \limits_{l = 1}^{L_\text{p}}  {\mathbf{P}_{q}^\text{H}[m,l]  \mathbf{P}_{q}[l,k] },   \quad   m,k = 1, ... , L_\text{p} \nonumber \\
&{}={}&  \sum \limits_{l = 0}^{L_\text{p}-1}  {{\mathbf{P}^\ast}_{q}[l,m] \mathbf{P}_{q}[l,k] },    \nonumber \\
&{}={}&  \sum \limits_{l = 0}^{L_\text{p}-1} \Big({p}_0^\ast[(l-m) \text{ mod }  L_\text{p}]  \exp({-j \omega_q (l-1)}) \nonumber \\
&{}\times{}& {p}_0[(l-k) \text{ mod } L_\text{p}] \exp({j \omega_q (l-1)})\Big)\nonumber \\
&{}={}&  \!\!\small{\sum \limits_{l = 0}^{L_\text{p}-1} \!\!{p}_0^\ast[(l-m) \text{\! mod \!}  L_\text{p}] {p}_0[(l-k) \text{\! mod \!} L_\text{p}] }.
\end{IEEEeqnarray}}\!
{Thus, the matrix $\mathbf{P}_{q}^\text{H}\mathbf{P}_{q}$ is only dependent on pilot symbols.}
Considering the $\text{MSE}$ as a metric to evaluate the performance of different pilot sequences, an exhaustive search can be performed {offline} to find the pilot sequence denoted as $\textbf{p} = \big[p_0[0], p_0[1]\ldots, p_0[L_\text{p}-1]\big]^\text{T}$ that minimizes the $\text{MSE}$. We have
\begin{IEEEeqnarray}{rcl}\label{eqP2}
\textbf{p} &{}={}& \text{arg} \displaystyle \min_{\textbf{p} \in \mathcal{S}_\text{p}}  \biggl( \frac{\sigma_n^2}{ {L_\text{p}}.\sigma^2_\text{p}} \sum\limits_{q=0}^{Q}  \text{tr} \left( {(\mathbf{V}^\text{H}\mathbf{P}_{q}^\text{H}\mathbf{P}_{q}\mathbf{V})^{-1}} \right)\biggr).
\end{IEEEeqnarray}
Here, $\mathcal{S}_\text{p}$ represents the constellation set of pilot symbols.

It is important to highlight that, in designing the sub-block length (pilot length), the sub-block length  $L_\text{p} = L_\text{c} + L_\text{h} -1$ represents the minimum pilot sequence length required to capture the ISI of both channel and FTN signaling. On the one hand, our objective is to identify the pilot sequence that minimizes the MSE of the channel estimation. Using the shortest feasible pilot sequence simplifies the offline exhaustive search for determining the optimal pilot sequence. On the other hand, our channel estimation algorithm benefits from the property that the mean of the data, $s_{\text{d},b}[m]=s_{\text{d}}[k=mL_\text{p}+b]$, is zero in the channel estimation algorithm. 
Thus, when the sub-block length ($L_\text{p}$) is short, and thus, the number of sub-blocks in a frame, $N_\text{m}$, is greater, the resulting constructed sequence, $s_{\text{d},b}[m]$, for a given $b$ and $m = 0,1, ..., N_\text{m}$ is long, ensuring the mean of the data samples to approach zero more closely.
\subsection{SP-aided Channel Estimation Algorithm for FTN signaling}
Our proposed SP-aided algorithm to estimate doubly-selective channels for FTN signaling is summarized as:
%%%%%%%%%%%%%%%%%%%%%%%%%%%%%%%%%%
\begin{algorithm}
  \caption{SP-aided Channel Estimation Algorithm for FTN signaling}
  \textbf{Input}:  $\tilde{r}[k]$ for $k=0,1, ..., N-1$, $L_\text{p}$ ,$Q$ , $\omega_q$, $\sigma_\text{p}$, and  $(\mathbf{P}_{q}  \mathbf{V} )^{-1}$ for $q=0,1, ..., Q$. \\
1) \textbf{For} $b=0, 1, ..., L_\text{p}-1$ \textbf{do} \\
       \hspace*{2.5em} $\bullet$ $\tilde{r}_b[m] = \tilde{r}_b[mL_\text{p}+b]$, $m = 0,1, ..., N_\text{m}$. \\
       \hspace*{2.5em} $\bullet$ Calculate $\tilde{r}^f_b(\xi)$ from (\ref{equation15}). \\
    \hspace*{1em} \textbf{EndFor}\\
  2) \textbf{For}  $q=0,1, ..., Q$ \textbf{do} \\
       \hspace*{2.5em} $\bullet$ Calculate $\tilde{r}^f_b(\xi)$ at $\xi = L_\text{p} \omega_q$ for $b=0, 1, ..., L_\text{p}-1$.\\
       \hspace*{2.5em} $\bullet$ $\tilde{\mathbf{r}}^f_q=[\tilde{r}^f_0(L_\text{p} \omega_q ), \tilde{r}^f_1(L_\text{p} \omega_q ), \hdots \tilde{r}^f_{L_\text{p}-1}(L_\text{p} \omega_q )]^\text{T}$. \\
       \hspace*{2.5em} $\bullet$ Calculate ${\bm{\hat{\lambda}}}_q$ from (\ref{equS5wh}). \\
    \hspace*{1em} \textbf{EndFor}\\ 
3) \textbf{For} $k=0, 1, ..., N-1$ \textbf{do} \\
        \hspace*{2.5em}  $\bullet$ Calculate $\mathbf{\hat{c}}_k = \big[\hat{c}_0[k], \hat{c}_1[k], \hdots, \hat{c}_{L_\text{c}-1}[k]\big]^\text{T}$ from \\
        \hspace*{3.5em}(\ref{equation9}). \\
    \hspace*{1em} \textbf{EndFor}\\
  \textbf{Output}:  ${\bm{\hat{\lambda}}}_q$ for $q=0,1, ..., Q$ and $\mathbf{\hat{c}}_k$ for $k=0, 1, ..., N-1$. 
  \label{alg:MYALG}
\end{algorithm}
%%%%%%%%%%%%%%%%%%%%%%%%%%%%%%%%%%%

%%%%%%%%%%%%%%%%%%%%%%%%%%%%%%%%%%%
\begin{algorithm}
  \caption{SP-aided SCEDD Algorithm for FTN signaling}
  \textbf{Input}:  $\tilde{r}[k]$ for $k=0,1, ..., N-1$, $L_\text{p}$ ,$Q$ , $\omega_q$, $\sigma_\text{p}$, and  $(\mathbf{P}_{q}  \mathbf{V} )^{-1}$ for $q=0,1, ..., Q$. \\
1) Finding $\mathbf{\hat{c}}_k$ for $k=0, 1, ..., N-1$ from Algorithm \ref{alg:MYALG}.\\
2) Estimating $\hat{\textbf{a}}$ employing the turbo equlizer.\\
  \textbf{Output}:  $\hat{\textbf{a}}$. 
  \label{alg:MYALG2}
\end{algorithm}
%%%%%%%%%%%%%%%%%%%%%%%%%%%%%%%%%%%%%%%%%%%%%%%%%
\begin{table*}[t]
  \centering
  \caption{{Complexity comparison of the PSLI algorithm and channel estimation in this paper with the works in \cite{ishihara2017iterative} and \cite{keykhosravi2023pilot}.} }
  \begin{tabular}{|c|c|}
    \hline
    Algorithm & Computational complexity \\
    \hline
    MP-aided method in \cite{ishihara2017iterative} &  $\mathcal{O}\Big({N_\text{p} \log}(N_\text{p}) + N_\text{p} +Itr \cdot N \cdot (L_\text{f}^2 +L_\text{eff}^2) +Itr \cdot N  \cdot2^K\Big)$ \\
    \hline
    MP-aided method in \cite{keykhosravi2023pilot}&
    $\mathcal{O}\Big( (N_\text{p}-L_\text{h}-L_\text{c}+1) \cdot (L_\text{c}+1) +Itr \cdot N \cdot (L_\text{f}^2 +L_\text{eff}^2) +Itr \cdot N  \cdot2^K\Big)$\\
    \hline
    Proposed SP-aided SCEDD method 
    &$\mathcal{O}\big(Q \cdot L_\text{p} \cdot L_\text{c} +N \cdot Q \cdot L_\text{c} + Itr \cdot N \cdot (L_\text{f}^2 +L_\text{eff}^2)+Itr \cdot N  \cdot2^K\big) $\\
\hline
    Proposed SP-aided JCEDD method 
    & $\mathcal{O}\big(Itr \cdot Q \cdot L_\text{p} \cdot L_\text{c} + Itr \cdot N \cdot Q \cdot L_\text{c} + Itr \cdot N \cdot (L_\text{f}^2 +L_\text{eff}^2)+Itr \cdot N  \cdot2^K \big)$
  \\
    \hline
    \end{tabular}
  \label{tab:table_2}
\end{table*}
\section{The SP-aided SCEDD and JCEDD methods For FTN Signaling}
\subsection{The SP-aided SCEDD Method For the FTN Signaling}
In this subsection, we present the SP-aided SCEDD method building upon our previously proposed SP-aided method for doubly-selective channel estimation for FTN signaling in Section ~\RomanNumeralCaps{3}. As illustrated in Fig. \ref{fig:SCEDD}, this algorithm performs channel estimation and data detection separately. The channel estimation process relies on the SP-aided technique outlined in Algorithm \ref{alg:MYALG} of Section ~\RomanNumeralCaps{3}. Data detection is carried out through a turbo equalizer. The steps of the proposed SP-aided SCEDD algorithm are summarized as shown in Algorithm \ref{alg:MYALG2}.
\subsection{The SP-aided JCEDD Method For the FTN Signaling}
To further improve both MSE and BER, in this subsection, we propose an SP-aided JCEDD method to estimate
doubly-selective channels and jointly detect data for the FTN signaling. The block diagram of our proposed SP-aided JCEDD method is depicted in Fig. \ref{fig:JCEDD}. The key in the proposed SP-aided JCEDD method is that each iteration of the {Turbo Equalizer block} incorporates 
the latest channel estimation update, which is vital for improving the accuracy of equalization and decoding in the next step. 
It should be noted that the channel estimation is updated in the current iteration of the {Turbo Equalizer block} by utilizing the data bits estimated from the previous iteration of the {Turbo Equalizer block}.

As illustrated in Fig. \ref{fig:JCEDD}, the initial {Channel Estimation block} receives the $\tilde{r}[k]$ and, using the knowledge of $\sigma_\text{p}^2$, the power of the pilot in the transmitted signal, estimates the channel basis coefficients, ${\bm{\hat{\lambda}}}_q$, and channel coefficients, $\mathbf{\hat{c}}_k$, based on the algorithm \ref{alg:MYALG}. The channel basis coefficients, ${\bm{\hat{\lambda}}}_q$, are then fed into the Pilot Contribution Removal block. The block Pilot Contribution Removal also receives the $\tilde{r}[k]$ and $\sigma_\text{p}^2$. The block Pilot Contribution Removal first calculates the $\hat{u}[k]$ according to (\ref{equation92}) and then substitute it in (\ref{equation91}) to find the $\hat{\mu}[k]$. 

The {Turbo Equalizer block} in Fig. \ref{fig:JCEDD} is based on a {linear soft-input soft-output} (SISO) MMSE equalizer that receives the $\hat{\mu}[k]$ and $\mathbf{\hat{c}}_k$ to calculate the soft information on the data bits for the next iteration of the {Turbo Equalizer block} and output an estimate on the data bits, $\hat{\textbf{a}}$. The estimated data bits output of the zero-th iteration of {the Turbo Equalizer block} is encoded, interleaved, and mapped to symbols and then fed into the Pilot Contribution Calculator block. The Pilot Contribution Calculator block also receives the  ${\bm{\hat{\lambda}}}_q$ and calculates the $\hat{u}[k]$ according to (\ref{equation92}) and feeds it into the {Channel Estimation block}. 

The {Channel Estimation block} in Fig. \ref{fig:JCEDD} receives the $\hat{u}[k]$ which is the pilot part of the $\tilde{r}[k]$ and thus it employs the algorithm \ref{alg:MYALG} for channel estimation having $\sigma_\text{p}^2 = 1$. The {Channel Estimation block} then outputs the updated ${\bm{\hat{\lambda}}}_q$ and feeds it into the Pilot Contribution Removal and Pilot Contribution Calculator blocks for the next iteration of {Turbo Equalizer block}. The {Channel Estimation block} also outputs the updated estimation of channel coefficients, $\mathbf{\hat{c}}_k$, into the {Turbo Equalizer block} for use in the next iteration. It should be noted that each iteration of the {Turbo Equalizer block} employs the updated estimation of channel coefficients, $\mathbf{\hat{c}}_k$, from {Channel Estimation block} and the updated $\hat{\mu}[k]$ from the Pilot Contribution Removal block as illustrated in Fig. \ref{fig:JCEDD}. The procedures of the proposed SP-aided JCEDD algorithm are outlined in Algorithm \ref{alg:MYALG3}.

%%%%%%%%%%%%%%%%%%%%%%%%%%%%%%%%%%%%%%%%%%%%%%%%%%%%%%%%%%%%%%%%%%

%%%%%%%%%%%%%%%%%%%%%%%%%%%%%%%%%%
\begin{algorithm}
  \caption{SP-aided JCEDD Algorithm for FTN signaling}
  \textbf{Input}:  $\tilde{r}[k]$ for $k=0,1, ..., N-1$, $L_\text{p}$ ,$Q$ , $\omega_q$, $Itr$, $\sigma_\text{p}$, and  $(\mathbf{P}_{q}  \mathbf{V} )^{-1}$ for $q=0,1, ..., Q$. \\
\textbf{Initialize}: Finding ${\bm{\hat{\lambda}}}_q$ for $q=0,1, ..., Q$ and $\mathbf{\hat{c}}_k$ for $k=0, 1, ..., N-1$ from Algorithm \ref{alg:MYALG}.\\
\textbf{For} $n_{Itr}=0, 1, ..., Itr-1$ of the Turbo equalizer \textbf{do}\\
\hspace*{1em}  1)  Pilot Contribution Removal: \\
        \hspace*{2.5em}  $\bullet$  Calculate $\hat{u}[k]$ from (\ref{equation92}). \\
        \hspace*{2.5em}  $\bullet$  Calculate $\hat{\mu}[k]$ from (\ref{equation91}).\\
\hspace*{1em}  2)  Turbo equalizer: \\
        \hspace*{2.5em}  $\bullet$  Calculate $\hat{\textbf{a}}$. \\
        \hspace*{2.5em}  $\bullet$  Calculate soft information on the bits.\\     
\hspace*{1em}  3)  Encoding \\
\hspace*{1em}  4)  Interleaving\\
\hspace*{1em}  5)  Mapping \\
\hspace*{1em}  6)  Pilot Contribution Calculator: \\
        \hspace*{2.5em}  $\bullet$  Calculate $\hat{u}[k]$ from (\ref{equation92}). \\    
\hspace*{1em}  7)  Channel Estimator: \\
        \hspace*{2.5em}  $\bullet$  Finding ${\bm{\hat{\lambda}}}_q$ for $q=0,1, ..., Q$ and $\mathbf{\hat{c}}_k$ for \\
        \hspace*{2.5em} $k=0, 1, ..., N-1$ from Algorithm \ref{alg:MYALG} for $\sigma_\text{p}=1$. \\        
        \textbf{EndFor}\\
\textbf{Output}:  $\hat{\textbf{a}}$  
\label{alg:MYALG3}
\end{algorithm}

{\subsection{Complexity Analysis}
This subsection presents the complexity analysis of the proposed SP-aided SCEDD and JCEDD algorithms and compares them with MP-aided methods in \cite{keykhosravi2023pilot} and \cite{ishihara2017iterative} as summarized in Table \ref{tab:table_2}.} 
{First we calculate the complexity of Algorithm 1 employed in SP-aided SCEDD and JCEDD algorithms. The first two steps of Algorithm 1 involve computing $\tilde{r}^f_b(\xi)$ from (\ref{equation15}) at $\xi = L_\text{p}\omega_q$ for $b=0, 1, ..., L_\text{p}-1$ and $q=0,1, ..., Q$, requiring $\mathcal{O}\big(Q\cdot  N \big)$ computational complexity where $N  = N_\text{m} \cdot {L_\text{p}}$. 
The complexity of computing of ${\bm{\hat{\lambda}}}_q$ from (\ref{equS5wh}) %for $q=0,1, ..., Q$ 
is $\mathcal{O}\big( Q \cdot L_\text{p} \cdot L_\text{c} \big)$ having the matrix inversion in (\ref{equS5wh}) precomputed offline at the receiver.
The third step of Algorithm 1 calculates $\mathbf{\hat{c}}_k$ from (\ref{equation9}) for $k=0, 1, ..., N-1$ which requires the complexity of $\mathcal{O}\big( N \cdot Q \cdot L_\text{c} \big)$. Thus the total complexity of Algorithm 1 is $\mathcal{O}\big(Q \cdot L_\text{p} \cdot L_\text{c} +N \cdot Q \cdot L_\text{c}\big)$.}

{The SP-aided SCEDD Algorithm in Algorithm 2 begins by executing Algorithm 1, followed by a turbo equalizer employing a SISO MMSE approach with $Itr$ iterations. As shown in \cite{tuchler2002turbo}, its per-iteration complexity is $\mathcal{O}\big(N \cdot (L_\text{f}^2 +L_\text{eff}^2)\big)$, where $L_\text{f}$ and $L_\text{eff}$ represent linear filter and the effective ISI length, respectively.  
The decoder in each iteration involves a computational complexity of $\mathcal{O}\big(N  \cdot2^K\big)$ where $K$ is the memory of the convolutional code. The deinterleaver applies a predefined permutation to the received code bits, operating with linear complexity.
Consequently, the total computational complexity of Algorithm 2 is $\mathcal{O}\big(Q \cdot L_\text{p} \cdot L_\text{c} +N \cdot Q \cdot L_\text{c} + Itr \cdot N \cdot (L_\text{f}^2 +L_\text{eff}^2)+Itr \cdot N  \cdot2^K\big)$, where the lower-order term $N$ is omitted as it is dominated by higher-order terms.}

{The first step of the loop in Algorithm 3 involves computing $\hat{u}[k]$ from (\ref{equation92}) and $\hat{\mu}[k]$ from (\ref{equation91}), resulting in a computational complexity of $\mathcal{O}\big(Q \cdot  N \cdot L_\text{c} \big)$. The second step performs one iteration of the turbo equalizer and decoding, which has a computational complexity of $\mathcal{O}\big(N \cdot (L_\text{f}^2 +L_\text{eff}^2)+N  \cdot2^K\big)$. The third step involves convolutional encoding with an small and fixed constraint length, implemented using a finite-length shift register with a linear complexity. The complexity of interleaver in the fourth step is also linear. The fifth step involves symbol mapping, which groups the permuted bits into symbols based on a predefined constellation, resulting in $\mathcal{O}\big(N \big)$ complexity.
The step six requires a computational complexity of $\mathcal{O}\big(Q \cdot  N \cdot L_\text{c} \big)$. Finally, considering the execution of Algorithm 1, the total computational complexity of Algorithm 3 is $\mathcal{O}\big(Itr \cdot Q \cdot L_\text{p} \cdot L_\text{c} +Itr \cdot N \cdot Q \cdot L_\text{c} + Itr \cdot N \cdot (L_\text{f}^2 +L_\text{eff}^2)+Itr \cdot N  \cdot2^K \big)$.}%  where the lower-order term $N$ is omitted as it is dominated asymptotically.}

%%%%%%%%%%%%%%%%%%%%%%%%%%%%%%%%%%%%%%%%%%%%%%%%%%%%%%%%%%%%
\section{Simulation Results}
This section offers a detailed analysis of the MSE and BER performance, providing numerical results to evaluate the effectiveness of our proposed SP-aided SCEDD and SP-aided JCEDD methods for FTN signaling proposed in Section ~\RomanNumeralCaps{4}. It should be noted that both SP-aided SCEDD and SP-aided JCEDD methods employs SP-aided doubly-selective channel estimation method proposed in Section ~\RomanNumeralCaps{3} for estimating doubly-selective channel in FTN signaling.

\subsection{Simulation Setup}
At the transmitter, a symbol rate of 2400  symbols/sec is adopted. A convolutional code with a rate of $1/2$ and a constraint length of 7 is utilized, characterized by the generator polynomials 0x5b and 0x79. By applying a puncturing pattern of $[1, 1, 1, 0, 0, 1]$ to this code, we achieve a code rate of $R_\text{c} = 3/4$, where a “1” indicates that the bit is transmitted, and a “0” signifies that the bit is omitted. At the receiver, decoding is performed using an a posteriori probability (APP) algorithm. Binary phase shift keying (BPSK) is used for pilot symbols, while quadrature phase shift keying (QPSK) is employed for data symbols. An exhaustive search was conducted to determine the pilot sequence, $\textbf{p}$, that minimizes the MSE, as described in (\ref{equS6wh1}). The optimal pilot sequence is identified through the application of (\ref{eqP2}).

To evaluate the MSE simulation results, the MSE for each simulation run is computed using the $\sum\nolimits_{l=0}^{L_{c}} ({\hat{\textbf{c}}_l}-\textbf{c}_l)^\text{H}({\hat{\textbf{c}}_l}-\textbf{c}_l)$. The final MSE is then obtained by averaging these MSE values over all simulation runs.
It is worth mentioning that, we employ the basis expansion model of (\ref{equation9}) with $\omega_q = 2 \pi (q-Q/2)/N$, $Q=2 \lceil{f_dN\tau T} \rceil$ to represent the the doubly-selective fading channel as described earlier. When modeling the doubly-selective fading channel with a finite basis, we introduce an expansion MSE (the MSE of the Basis Expansion Model), which results in a channel modeling error reflected in the MSE of channel estimation. In other words, the MSE of channel estimation that we calculate in the simulation arises from two sources of errors: (1) the errors in estimating the basis coefficients, $\lambda_{q}[l]$, and (2) the basis expansion errors, which stems from constructing the channel coefficients, $c_l[k]$, from the finite basis coefficients. We define the $\alpha = \sigma_\text{p}^2/(\sigma_\text{d}^2+\sigma_\text{p}^2)$ as the power allocation factor between the pilot and data. In our simulations, the signal-to-noise ratio is defined as $(\sigma_\text{d}^2+\sigma_\text{p}^2)/\sigma_n^2$ where $\sigma_n^2$ is the power of the zero-mean white Gaussian noise.

For the simulation in this paper, we focus on the HF channel model introduced by the International Telecommunications Union Radio Communication Sector (ITU-R). A comprehensive set of HF channel models is recommended by the ITU-R and is widely used in the performance assessment of HF systems. Specifically, ITU-R F.1487 \cite{otnes2002improved} introduces a comprehensive set of 10 test channels, encompassing various latitude regions and levels of ionospheric disturbance. 
The ITU-R F.1487 recommendations outline all test channels as tapped delay line models featuring only two non-zero taps. These taps experience independent fading, following a Rayleigh probability density function. Additionally, both paths experience equal average power with the same Doppler frequency.
We employ two practical channel models for simulations in this study to consider the effects of Doppler frequency on the results: 1) The high-latitude channel, which comprises two non-zero independent paths with a fixed delay spread of 3 ms and a Doppler frequency of 10 Hz (fading rate of $0.004$). This channel model will be denoted as \textit{channel model 1}.
2) ITU-R Poor channel (low-latitude channel) which consists of two non-zero independent paths with a fixed delay spread of 2.1 ms and a Doppler frequency of 1 Hz (fading rate of $0.0004$). This channel model will be denoted as \textit{channel model 2}. This channel model is employed in \cite{keykhosravi2023pilot}.

The SE measured in bits/sec/Hz in an MP-aided FTN signaling system and our proposed SP-aided algorithms for FTN signaling system is defined respectively as $\gamma_{\text{MP-FTN}}$, and $\gamma_{\text{SP-FTN}}$ 
\begin{IEEEeqnarray}{rcl}\label{equationSE}
\gamma_{\text{MP-FTN}} &{}={}&  \frac{{{N_\text{s}}}}{({{N_\text{s}}}+N_\text{p})} \cdot\frac{{\log_2}M}{\tau (1+\beta)}R_\text{c},\\
\gamma_{\text{SP-FTN}} &{}={}&   \frac{{\log_2}M}{\tau (1+\beta)}R_\text{c},
\end{IEEEeqnarray}
where $\beta$ is the roll-off factor of the RRC pulse shaping filter and $R_\text{c}$ is the code rate. {Unless otherwise specified, our simulations employ a roll-off factor of $\beta = 0.35$, which is typical for HF communication. We also compare the system’s performance with a lower roll-off factor of $\beta = 0.22$.}
The $N_\text{s}$ and $N_\text{p}$ are the number of data symbols {and the overhead length (e.g., pilot, cyclic prefix, etc.)} within a frame in an MP-aided FTN signaling system. If the systems use $M$-ary quadrature amplitude modulation ($M$-QAM) constellation set and the same $\beta$ and $R_\text{c}$, the {values of $\tau$, $N_\text{s}$ and $N_\text{p}$ must be carefully chosen to ensure a fair SE comparison} between the systems. In the MP-aided case presented in \cite{keykhosravi2023pilot} {and \cite{ishihara2017iterative}}, the pilot sequence duration must be carefully designed to be shorter than the channel coherence time, ensuring that the channel coefficients remain constant throughout this period, {while also exceeding the effective ISI, including both channel ISI and FTN induced ISI, for reliable channel estimation.}
{If we define the coherence time as the duration for which the autocorrelation function remains above 0.5, it is approximately $T_\text{c} = \frac{9}{16 \pi}$ as in \cite{rappaport2002wireless}.
Considering a transmission rate of 2400 symbol/sec over \textit{channel model 1}, with a Doppler frequency of 10 Hz, the channel remains coherent for approximately $N_\text{c} = 42$ symbols. We set $N_\text{p} = 32$ as a trade-off as choosing a lower $N_\text{p}$ degrades channel estimation quality due to high effective ISI in FTN signaling transmission, whereas a higher $N_\text{p}$ would exceed the coherence time. {This constraint limits the MP-aided method’s performance as the fading rate increases, as in \textit{channel model 1}. Thus, at a fixed SE, these MP-aided reference methods are expected to underperform compared to our proposed SP-aided method due to severe ISI and low coherence time.} 
For \textit{channel model 2} which has a Doppler frequency of 1 Hz and a higher coherence time ($N_\text{c} = 429$ symbols), the pilot sequence length is set to $N_\text{p} = 32$, following \cite{keykhosravi2023pilot}. In this scenario, the longer coherence time and reduced impact of effective ISI contribute to a more reliable channel estimation process in low-fading-rate conditions like \textit{channel model 2}. }

{In the simulation setup for the MP-aided system, we ensure a fair comparison in terms of SE by considering two scenarios over \textit{channel model 1}: \textit{case A}) $\tau = 0.72$ with $N_\text{s} = 128$, and \textit{case B}) $\tau = 0.8$ with $N_\text{s} = 256$. We set $N_\text{p} = 32$ in \cite{keykhosravi2023pilot} resulting in an SE of 1.23 bits/s/Hz for both cases. Additionally, we consider $N_\text{p} = 40$ and $N_\text{p} = 44$ in \cite{ishihara2017iterative}, yielding SE of 1.20 bits/s/Hz and 1.14 bits/s/Hz for \textit{case B} and \textit{case A}, respectively. For our SP-aided approach, we set $\tau = 0.9$, achieving an SE of $1.23$ bits/s/Hz.
This setup facilitates a meaningful comparison between $\gamma_{\text{MP-FTN}}$ and $\gamma_{\text{SP-FTN}}$, where different strategies are employed to achieve similar SE: reducing $\tau$ or increasing $N_\text{s}$. A lower $\tau$ introduces stronger FTN-induced ISI, degrading both MSE and BER performance. On the other hand, increasing $N_\text{s}$ in MP-aided channel estimation leads to higher interpolation errors, negatively impacting MSE and BER. If $N_\text{s}$ approaches or exceeds the channel coherence time, these accumulated interpolation errors become more severe. In our setup, selecting $\tau=0.72$ (slightly lower than $\frac{1}{1+\beta} = 0.74$) and 0.8 ensures the multipath delays correspond to exactly 11 and 10 symbol intervals, respectively. {In \cite{ishihara2017iterative}, the overhead is four times the pilot sequence length. Thus, a pilot length of 10 symbols corresponds to an overhead of $N_\text{p} = 40$ symbols per frame.}}
{Extending the above methodology to \textit{channel model 2}, we set $N_\text{p} = 32$ in \cite{keykhosravi2023pilot} and \cite{ishihara2017iterative}. To maintain a fair SE comparison, we examine two scenarios for MP-aided transmission: \textit{case 1}) $\tau = 0.72$ with $N_\text{s} = 256$, achieving an SE of 1.37 bits/s/Hz, and \textit{case 2}) $\tau = 0.84$ with $N_\text{s} = 1024$, yielding the SE of 1.28 bits/s/Hz. For our SP-aided approach, we set $\tau = 0.84$, resulting in an SE of $1.32$ bits/s/Hz. 
Furthermore, for \textit{channel model 2}, selecting $\tau = 0.72$ and $\tau = 0.84$ ensures that the multipath delays correspond to exactly 8 and 7 symbol intervals, respectively.}

%%%%%%%%%%%%%%%%%%%%%%%%
\begin{figure}[t]
\centering
\includegraphics[width=8.35cm,height=5.62cm]{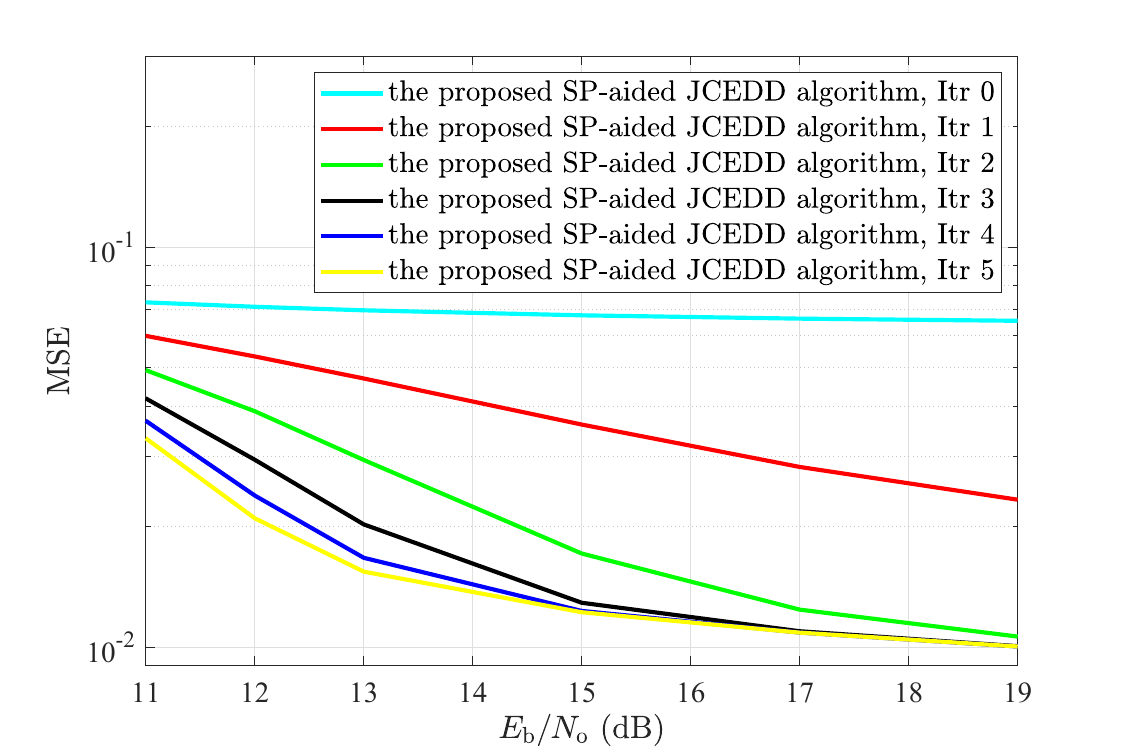} 
\caption{The MSE of proposed SP-aided JCEDD algorithm for FTN signaling over \textit{channel model 1} having $\tau = 0.9$ and $\alpha = 31\%$.}
\label{fig:21}
\end{figure}
%%%%%%%%%%%%%%%%%%%%%%%%
\begin{figure}[t]
\centering
\includegraphics[width=8.35cm,height=5.62cm]{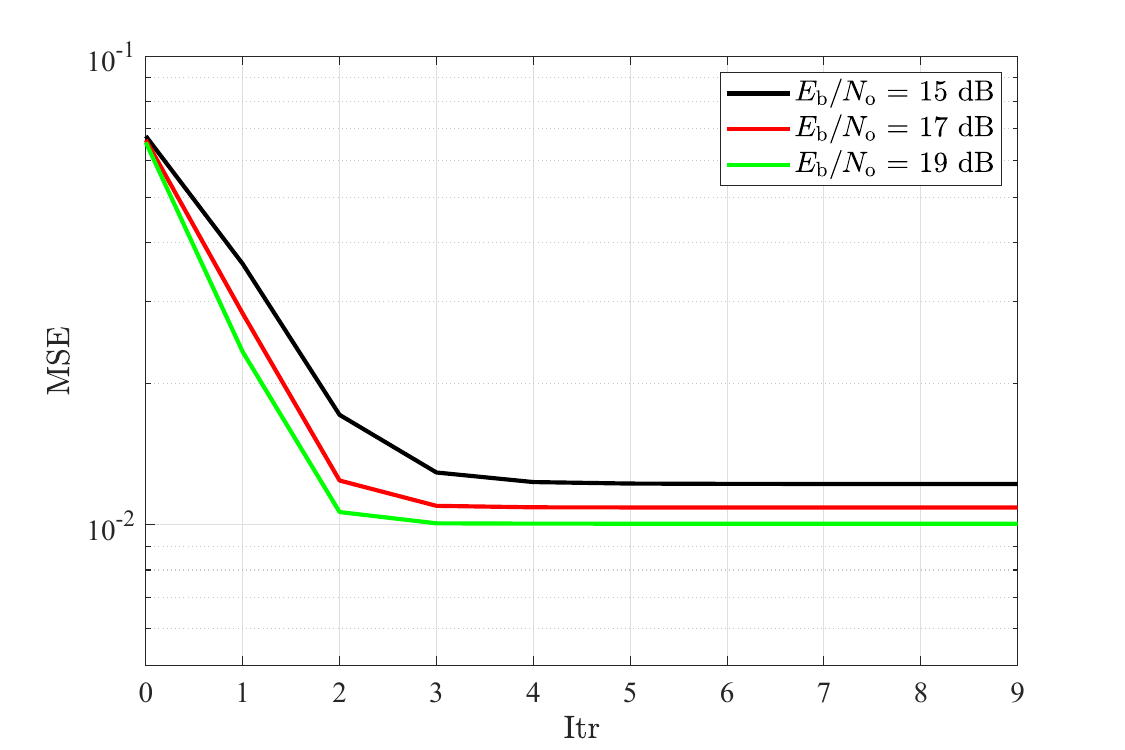} 
\caption{{The MSE of proposed SP-aided JCEDD algorithm for FTN signaling over \textit{channel model 1} having $\tau = 0.9$ and $\alpha = 31\%$.}} 
\label{fig:21it}
\end{figure}

%%%%%%%%%%%%%%%%%%%%%%%%%%%%%%%%%%%
\begin{figure}[t]
\centering
\includegraphics[width=8.35cm,height=5.62cm]{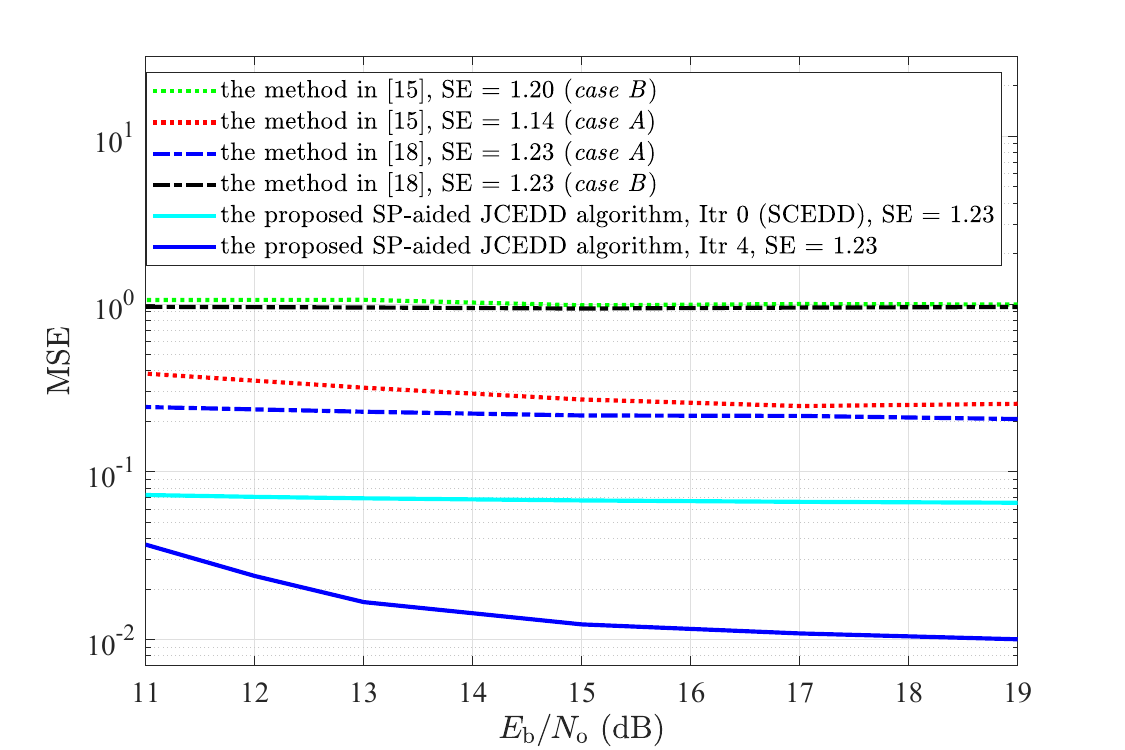} 
\caption{{The MSE of proposed SP-aided algorithms for FTN signaling  over \textit{channel model 1} having $\tau = 0.9$ and $\alpha = 31\%$.}}
\label{fig:22}
\end{figure}

%%%%%%%%%%%%%%%%%%%%%%%%%

\begin{figure}[t]
\centering
\includegraphics[width=8.35cm,height=5.62cm]{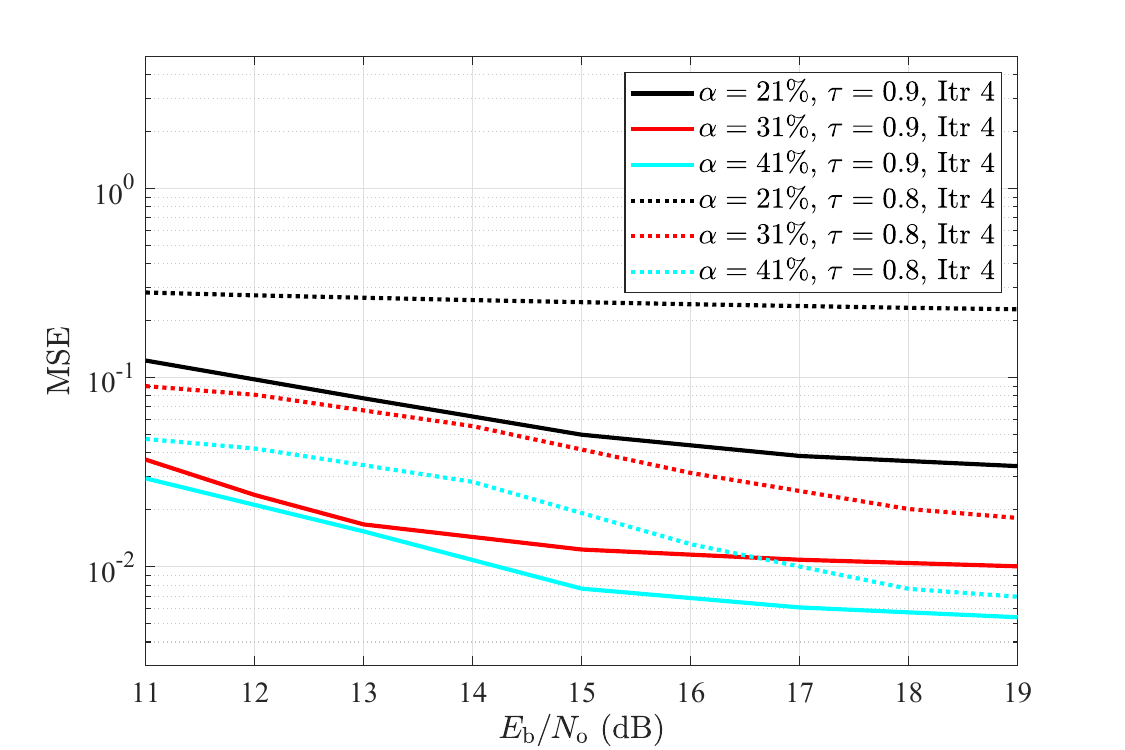} 
\caption{The MSE of proposed SP-aided JCEDD algorithm for FTN signaling over \textit{channel model 1}.}
\label{fig:23}
\end{figure}
%%%%%%%%%%%%%%%%%%%%%%%%%
\begin{figure}[t]
\centering
\includegraphics[width=8.35cm,height=5.62cm]{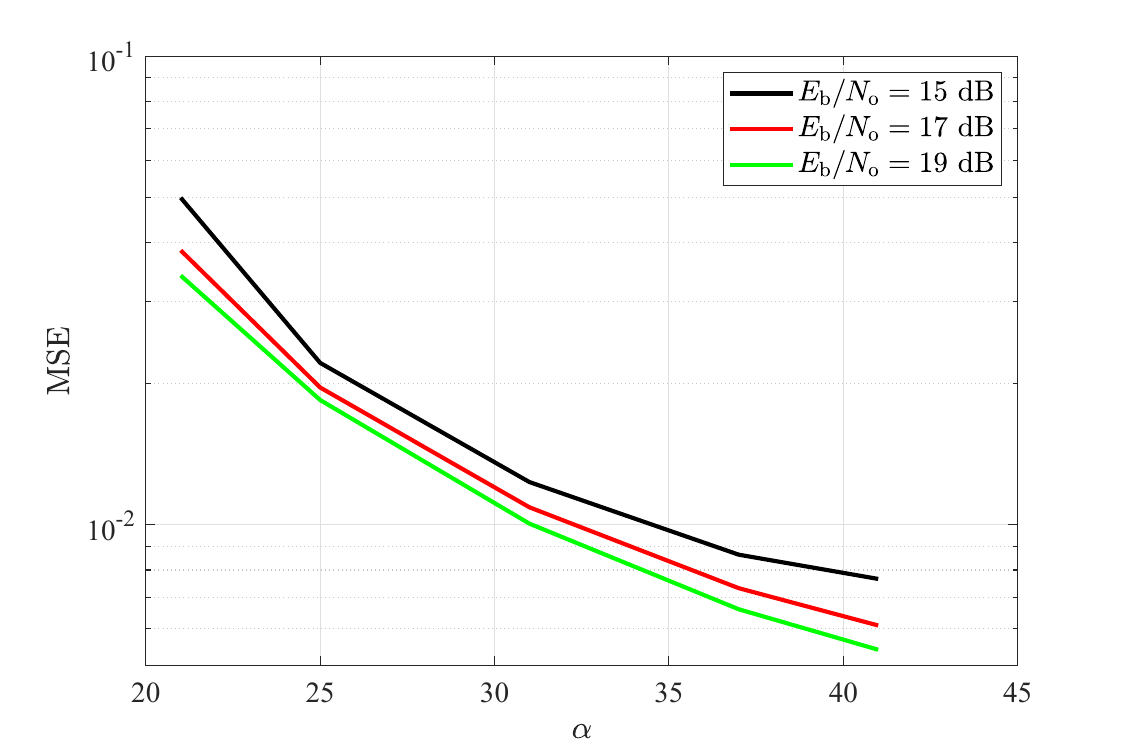} 
\caption{{The MSE of proposed SP-aided JCEDD algorithm for FTN signaling over \textit{channel model 1} having $\tau = 0.9$.}}
\label{fig:233}
\end{figure}

%%%%%%%%%%%%%%%%%%%%%%%%%

%%%%%%%%%%%%%%%%%%%%%%%%%%%%%%%%%%%%
\begin{figure}[t]
\centering
\includegraphics[width=8.35cm,height=5.62cm]{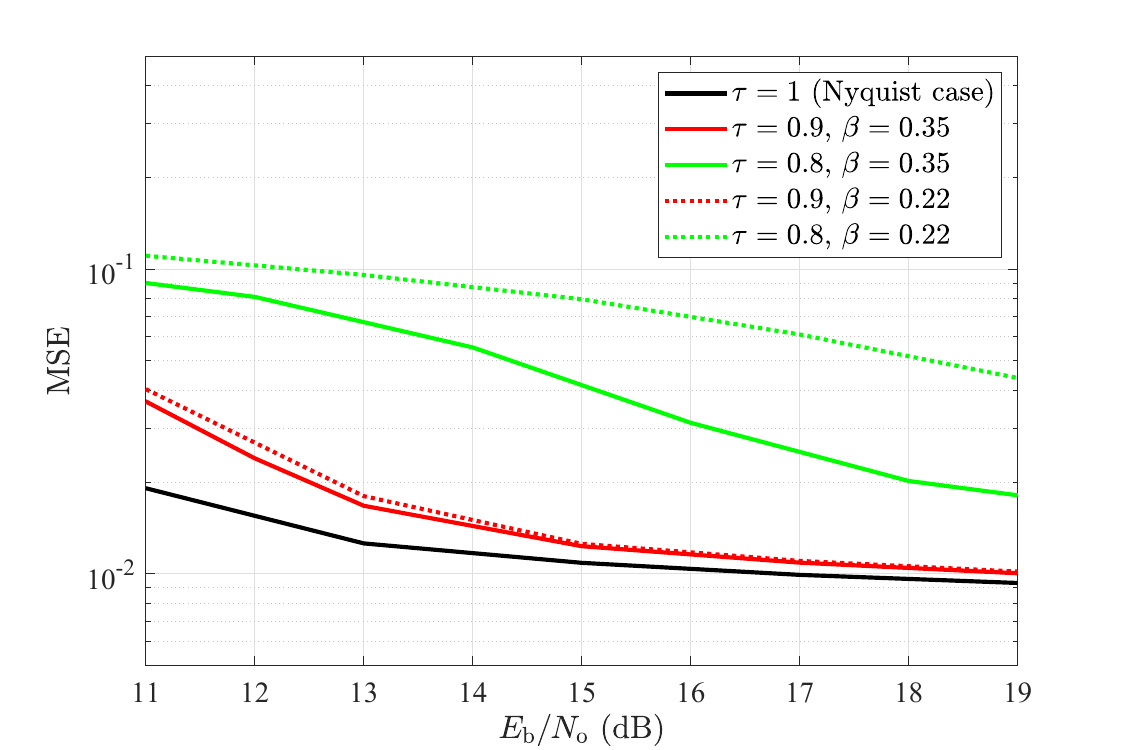} 
\caption{{The MSE of proposed SP-aided JCEDD algorithm for FTN signaling over \textit{channel model 1} with $\alpha = 31\%$.}}
\label{fig:25}
\end{figure}
%%%%%%%%%%%%%%%%%%%%%%%%%%%%%%%%%%%%

%%%%%%%%%%%%%%%%%%%%%%%%%%%%%%%%%%%
\begin{figure}[t]
\centering
\includegraphics[width=8.35cm,height=5.62cm]{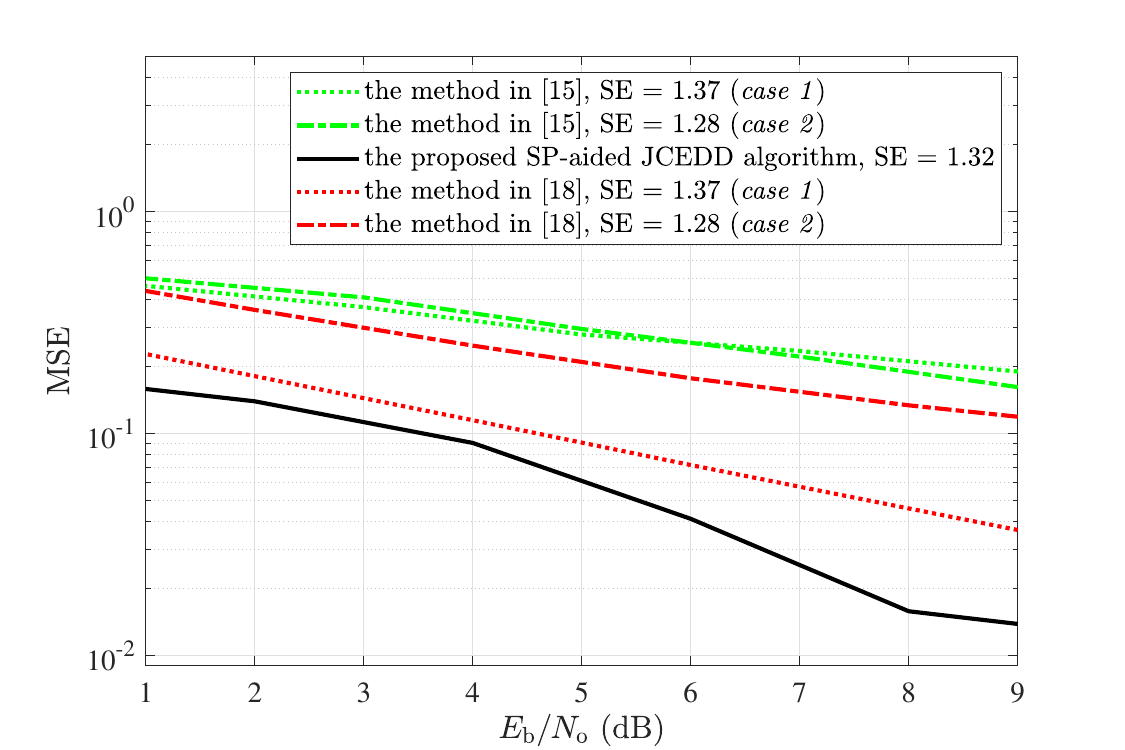} 
\caption{{The MSE of proposed SP-aided JCEDD algorithm for FTN signaling %with an SE of around $1.3$ bits/s/Hz 
over \textit{channel model 2} having $\alpha = 12\%$.} }
\label{fig:224}
\end{figure}
%%%%%%%%%%%%%%%%%%%%%%%%%%%%%%
\subsection{MSE Results}

%%%%%%%%%%%%%%%%%%%%%%%%%%%%%%%%%%%%%%
{Fig. \ref{fig:21} and Fig. \ref{fig:21it} illustrate the MSE of channel estimation for our proposed SP-aided JCEDD algorithm, utilizing a power allocation factor $\alpha = 31\%$ for FTN signaling with $\tau = 0.9$ over \textit{channel model 1}. The SE, $\gamma_{\text{SP-FTN}}$, as defined in (\ref{equationSE}) to equal to $1.23$ bits/s/Hz. As shown in Fig. \ref{fig:21} and Fig. \ref{fig:21it}, the MSE improves greatly after only two iterations, and as $E_\text{b}/N_0$ increased to {19} dB, the MSE approached a value of $10^{-2}$. As four iterations are enough for the MSE to reach to its minimum, we use the results of iteration four for our discussion.}

Fig. \ref{fig:22} compares the MSE of channel estimation for our proposed SP-aided JCEDD and SCEDD algorithms having power allocation factor $\alpha = 31\%$ for FTN signaling with $\tau = 0.9$ over \textit{channel model 1} with the MP-aided {methods presented in \cite{keykhosravi2023pilot} and \cite{ishihara2017iterative}}. As the channel estimation is not updated in each iteration of turbo equalizer, the MSE of channel estimation for our proposed SP-aided SCEDD algorithm is equivalent to the $0$th {iteration} of our proposed SP-aided {JCEDD} algorithm. 
{To ensure a fair SE comparison, we examine two configurations, \textit{case A} and \textit{case B}, for MP-aided transmission over \textit{channel model 1}, as previously outlined.} 
As shown in Fig. \ref{fig:22}, our proposed SP-aided SCEDD algorithm for FTN signaling significantly outperforms the method in{\cite{keykhosravi2023pilot} and \cite{ishihara2017iterative} for both \textit{case A} and \textit{case B} in terms of MSE. Moreover, the MSE of our proposed SP-aided JCEDD algorithm for FTN signaling is far superior to all of the aforementioned methods}.

%%%%%%%%%%%%%%%%%%%%%%%%%%%%%%%%%%%%%%%%%%%%%%%%%%%%%%%%%%

{Fig. \ref{fig:23} shows a comparison of the MSE for our proposed SP-aided JCEDD method of FTN signaling with $\tau = 0.9$ and $\tau = 0.8$}, over \textit{channel model 1} with different power allocation factors. The number of iterations is set to four for the simulation shown in {Fig. \ref{fig:23}}. 
As expected, for a given value of $\tau$, the MSE improves as the power allocation factor, $\alpha$, increases. 
{Additionally, for a fixed value of power allocation factor, $\alpha$, the MSE is better for a larger value of $\tau$.}
It should be noted that for a given value of $\tau$, we need to choose the power allocation factor carefully to achieve a good BER. On the one hand, a higher value of the power allocation factor results in a lower MSE of the channel estimation. However, less power is left for data transmission, resulting in a higher BER. Also, if the BER is high in one iteration, it is not expected that the channel estimation will improve highly in the next iteration. On the other hand, a low value of the power allocation factor results in a high MSE of the channel estimation {which in turn is expected to adversely affect BER}. 

{The MSE versus power allocation factor, $\alpha$, for our proposed SP-aided JCEDD method of FTN signaling with $\tau = 0.9$, over \textit{channel model 1} for three values of different $E_\text{b}/N_0$ is illustrated in Fig. \ref{fig:233}. For the simulation in Fig. \ref{fig:233}, the number of iterations is considered as 4. Fig. \ref{fig:233} confirms that for a given value of $E_\text{b}/N_0$ the MSE shows improvement as the power allocation factor increases. This result is expected as allocating more power to the pilot and less to the data results in a better channel estimation quality. For a fixed value of power allocation factor, the MSE is better for a larger value of $E_\text{b}/N_0$ as expected.}

Fig. \ref{fig:25} compares the MSE for our proposed SP-aided JCEDD method for FTN signaling for $\tau = 0.9$, and $\tau = 1$ (Nyquist case) over \textit{channel model 1}{, considering two roll-off factor values: $\beta = 0.35$ and $\beta = 0.22$.} The number of iterations is set to four for the simulation shown in Fig. \ref{fig:25}. 
As expected, {for a fixed value of $\beta$,} the MSE improves as $\tau$ increases. 
For {$\beta = 0.35$ and $\tau = 0.9$, Fig. \ref{fig:25} reveals that as $E_\text{b}/N_0$ values exceeds 17 dB, the MSE for $\tau = 0.9$ approaches that of the $\tau = 1$ (Nyquist case).}
The SE for the traditional Nyquist case is $1.11$ bits/s/Hz{. Compared to Nyquist signaling, the proposed SP-aided JCEDD approach for FTN signaling achieves an $11\%$ improvement in SE.
In FTN signaling, we select the value of $\tau$ such that multipath delays align as exact multiples of the symbol interval. To ensure a fair comparison with the FTN signaling case, we approximate the delay spread of \textit{channel model 1} to 2.917 ms, ensuring that multipath components align with integer multiples of the symbol interval under Nyquist signaling. Under these idealized Nyquist conditions, where each multipath delay is an integer multiple of the symbol interval, variations in the roll-off factor $\beta$ do not impact MSE or BER performance. This is because the delayed versions of the signal arrive exactly at the designated sampling instances, preventing additional ISI.}
{As shown in Fig. \ref{fig:25}, for a given symbol packing ratio $\tau$ in FTN signaling, decreasing $\beta$ adversely affects MSE. This is because, as $\beta$ decreases, the time domain RRC pulse decays more slowly. Thus, ISI increases at the FTN signaling sampling points, worsening MSE even for a fixed $\tau$. 
It is also evident from Fig. \ref{fig:25} that reducing $\tau$ amplifies the negative impact of lowering $\beta$ on MSE. In particular, for $\tau = 0.9$  a decrease in $\beta$ from 0.35 to o.22 results in a minor MSE degradation, whereas for $\tau = 0.8$, the same reduction in $\beta$ results in over 3 dB loss in MSE for $E_\text{b}/N_0$ above 16 dB. The reason lies in the fact that in FTN signaling, reducing $\tau$ increases the overlap between adjacent pulses. Thus, the time domain pulse shape, and consequently $\beta$ become more critical factors in determining ISI levels. When $\tau$ is small, the sampling instances fall into the region where small changes in $\beta$, which affects the steepness and decay of the RRC pulse, can introduce a significant amount of ISI at each FTN signaling sampling point, leading to greater MSE degradation. For larger $\tau$, the sampling points are close to Nyquist signaling sampling points, and thus the same change in $\beta$ introduces less ISI. }

%%%%%%%%%%%%%%%%%%%%%%%%%%%%%%%%%%%%%%
%%%%%%%%%%%%%%%%%%%%%%%%%%%%%%%%%%%%%%
%%%%%%%%%%%%%%%%%%%%%%%%%%%%%%%%%%%%%%
Fig. \ref{fig:224} illustrates a comparison of the MSE performance in channel estimation between our proposed SP-aided JCEDD algorithm, with a power allocation factor of $\alpha = 12\%$ and the MP-aided {methods presented in\cite{keykhosravi2023pilot} and \cite{ishihara2017iterative} for FTN signaling over \textit{channel model 2}. 
To ensure a fair SE comparison, we consider two configurations, \textit{case 1} and \textit{case 2} for MP-aided transmission over \textit{channel model 2} as introduced earlier.} 
{Fig. \ref{fig:224} demonstrates that increasing $N_\text{s}$ (\textit{case 2}) in \cite{keykhosravi2023pilot} has a more detrimental effect on MSE performance than reducing $\tau$ (\textit{case 1}), even though the latter results in a slightly higher SE. 
As depicted in Fig. \ref{fig:224}, the SP-aided JCEDD algorithm for FTN signaling offers over 2 dB and 4 dB improvement in MSE performance for $E_\text{b}/N_0$ values higher than 5 dB compared to the \textit{case 1} and \textit{case 2} in \cite{keykhosravi2023pilot}, respectively. Additionally, Fig. \ref{fig:224} reveals that the SP-aided JCEDD algorithm for FTN signaling achieves over 6 dB improvement in MSE compared to the approach in \cite{ishihara2017iterative}.}

%%%%%%%%%%%%%%%%%%%%%%%%
\begin{figure}[t]
\centering
\includegraphics[width=8.35cm,height=5.62cm]{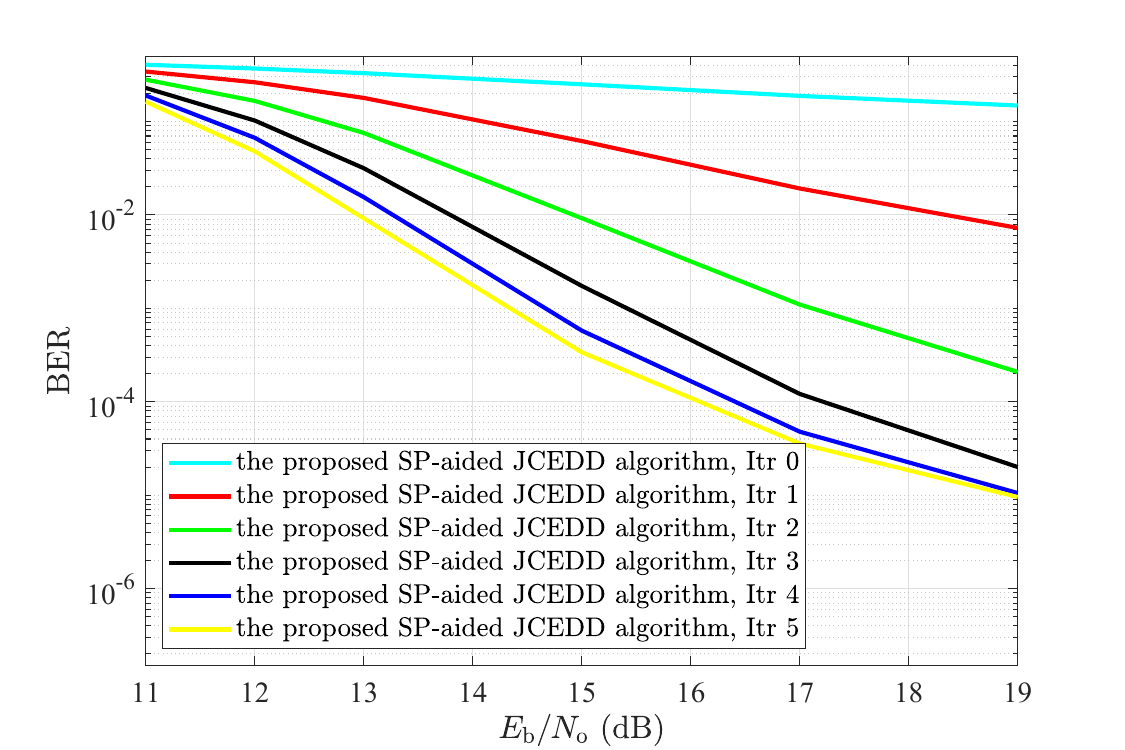} 
\caption{The BER performance of the proposed SP-aided JCEDD algorithm for FTN signaling over \textit{channel model 1} having $\tau = 0.9$ and $\alpha = 31\%$.}
\label{fig:121}
\end{figure}
%%%%%%%%%%%%%%%%%%%%%%%%
\begin{figure}[t]
\centering
\includegraphics[width=8.35cm,height=5.62cm]{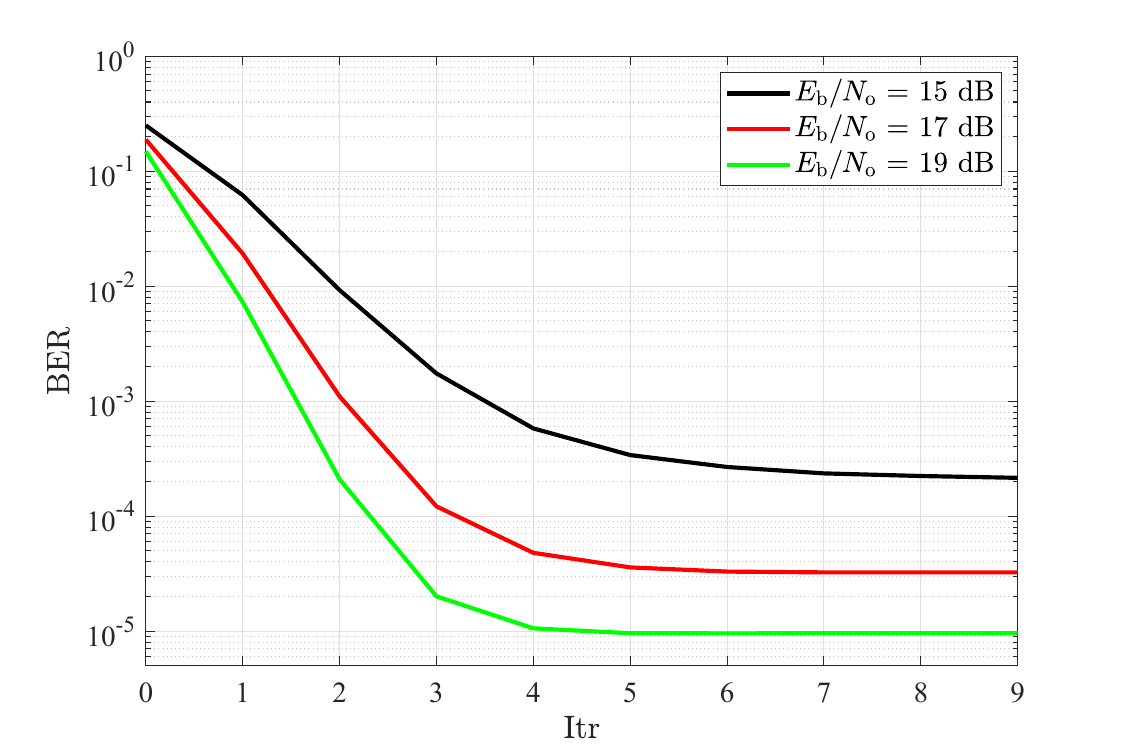} 
\caption{{The BER performance of the proposed SP-aided JCEDD algorithm for FTN signaling over \textit{channel model 1} with $\tau = 0.9$ and $\alpha = 31\%$.}} %for different values of $E_\text{b}/N_0$
\label{fig:121it}
\end{figure}
%%%%%%%%%%%%%%%%%%%%%%%%%%%%%%%%%%%
\begin{figure}[t]
\centering
\includegraphics[width=8.35cm,height=5.62cm]{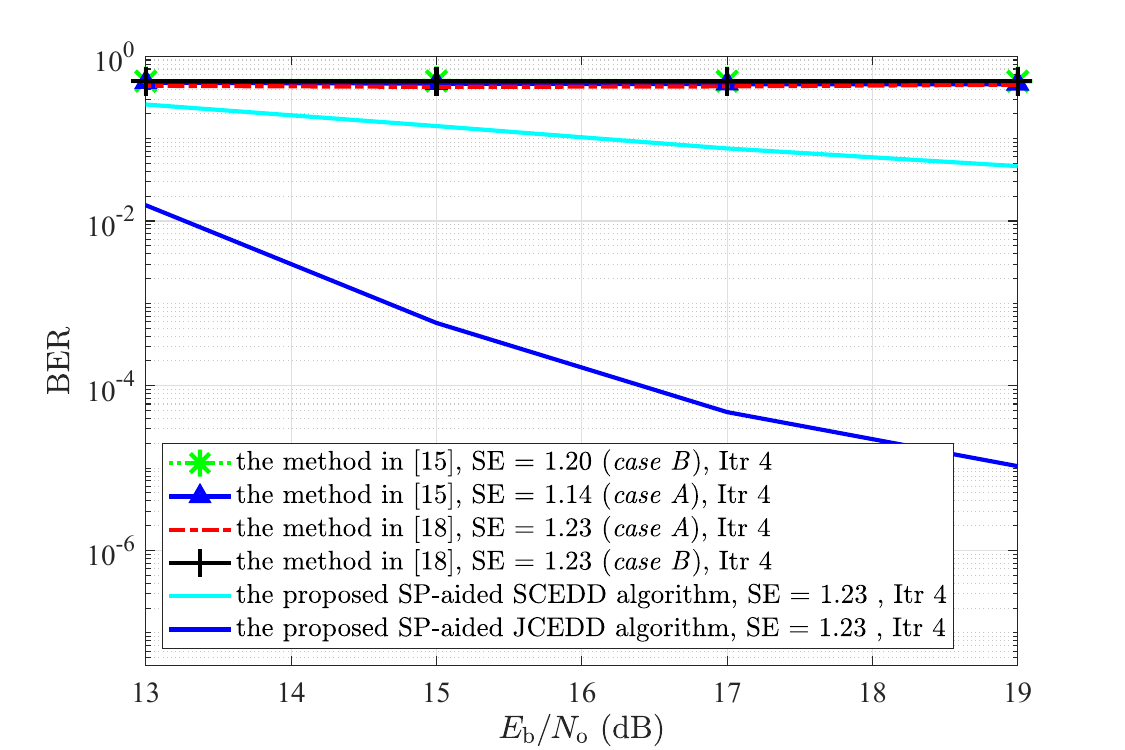} 
\caption{{The BER performance of the proposed SP-aided approaches for FTN signaling over \textit{channel model 1} having $\alpha = 31\%$.}}
\label{fig:122}
\end{figure}
%%%%%%%%%%%%%%%%%%%%%%%%%

\begin{figure}[t]
\centering
\includegraphics[width=8.35cm,height=5.62cm]{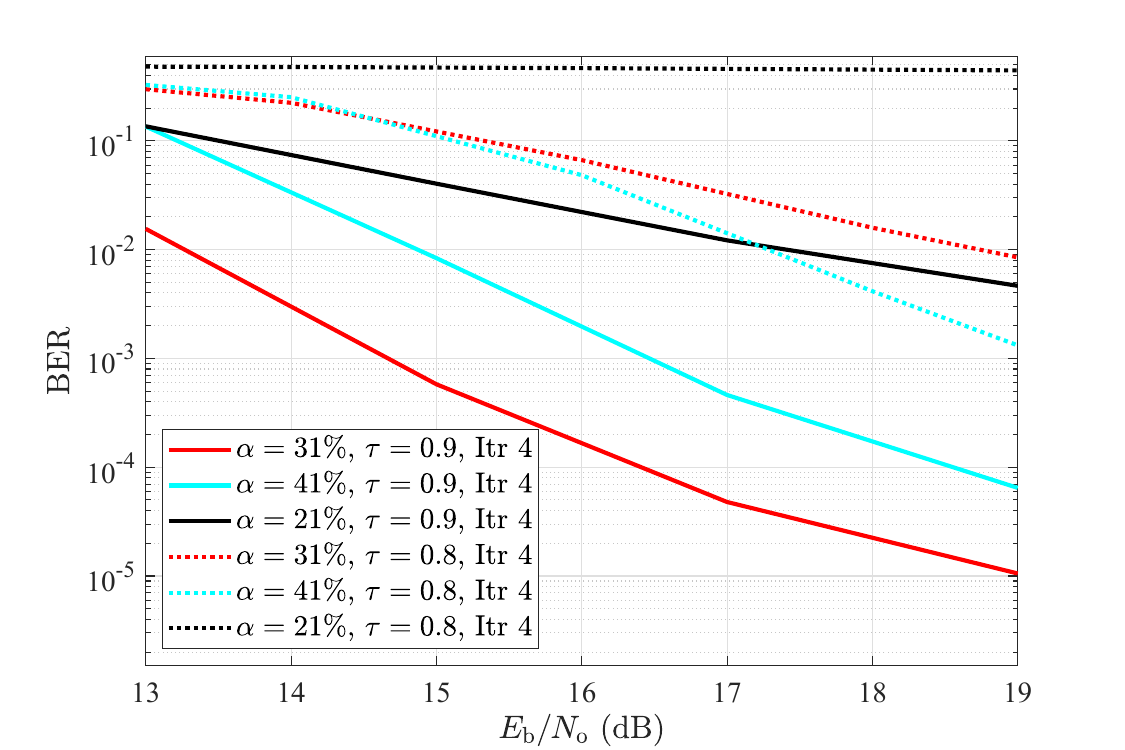} 
\caption{{The BER performance of the proposed SP-aided JCEDD approach for FTN signaling over \textit{channel model 1}.}}
\label{fig:123}
\end{figure}
%%%%%%%%%%%%%%%%%%%%%%%%%
\begin{figure}[t]
\centering
\includegraphics[width=8.35cm,height=5.62cm]{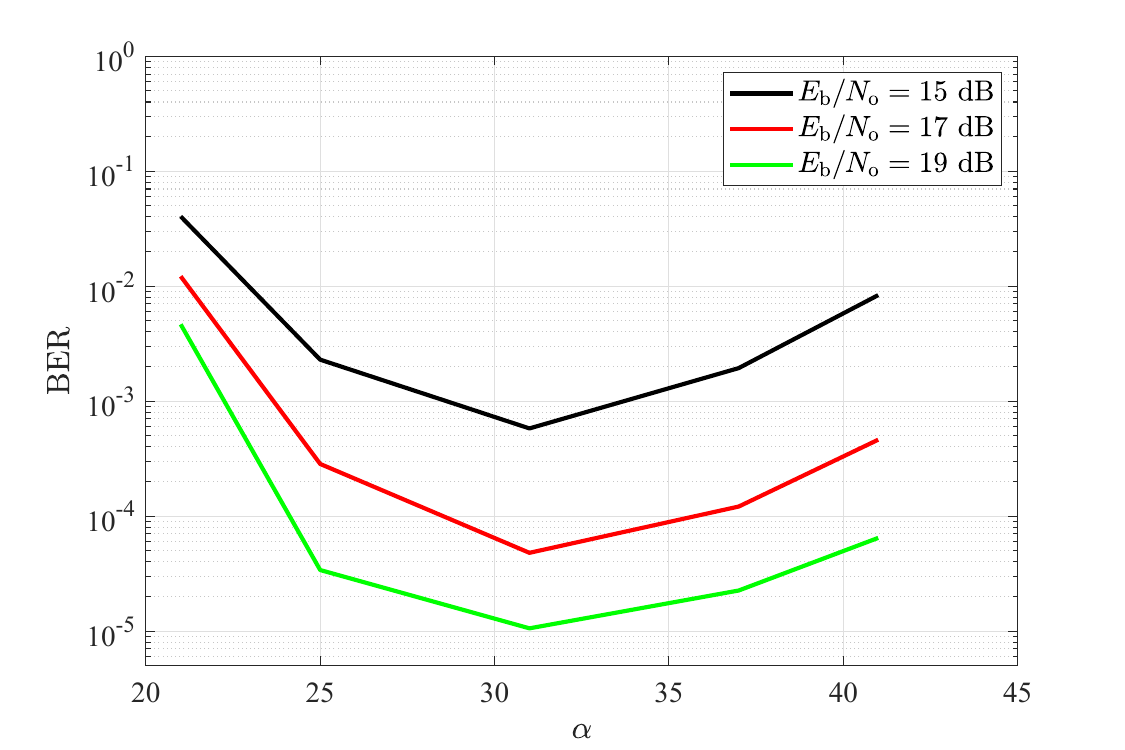} 
\caption{{The BER performance of the proposed SP-aided JCEDD approach for FTN signaling over \textit{channel model 1} having $\tau = 0.9$.}}
\label{fig:1234}
\end{figure}
%%%%%%%%%%%%%%%%%%%%%%%%%

%%%%%%%%%%%%%%%%%%%%%%%%%%%%%%%%%%%%
\begin{figure}[t]
\centering
\includegraphics[width=8.35cm,height=5.62cm]{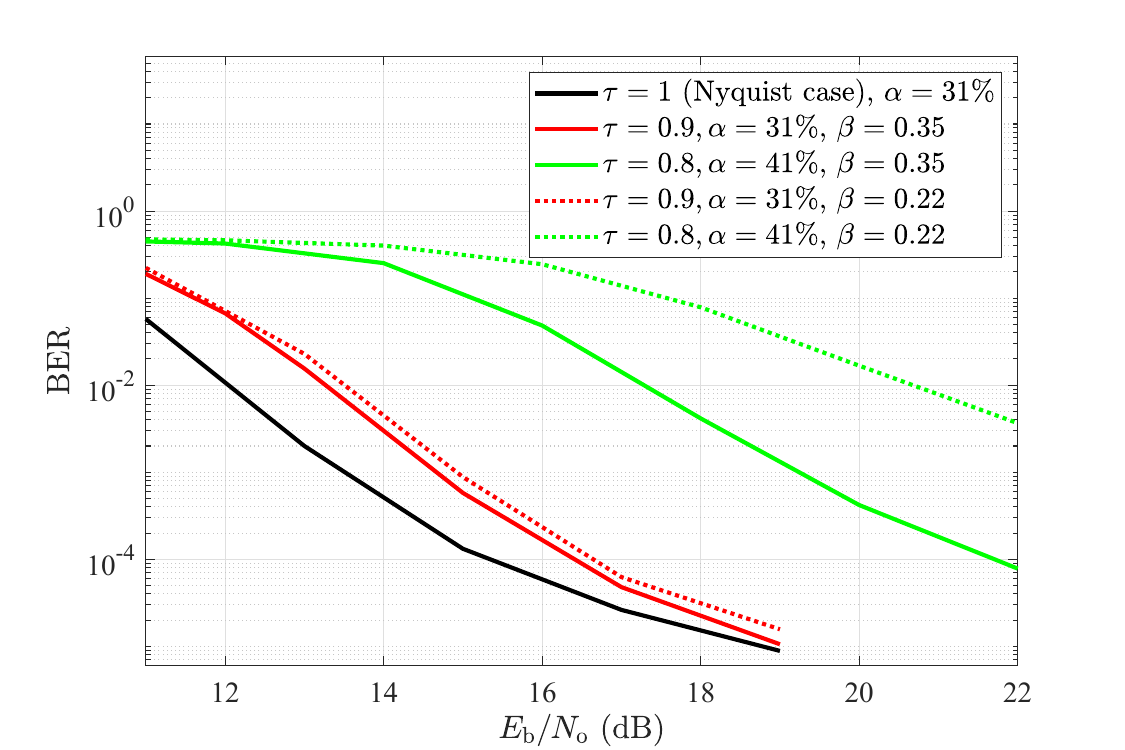} 
\caption{The BER performance of the proposed SP-aided JCEDD approach for FTN signaling over \textit{channel model 1}.}%{, with all methods simulated at their fourth iteration.}}
\label{fig:125}
\end{figure}
%%%%%%%%%%%%%%%%%%%%%%%%%%%%%%%%%%%
\begin{figure}[t]
\centering
\includegraphics[width=8.35cm,height=5.62cm]{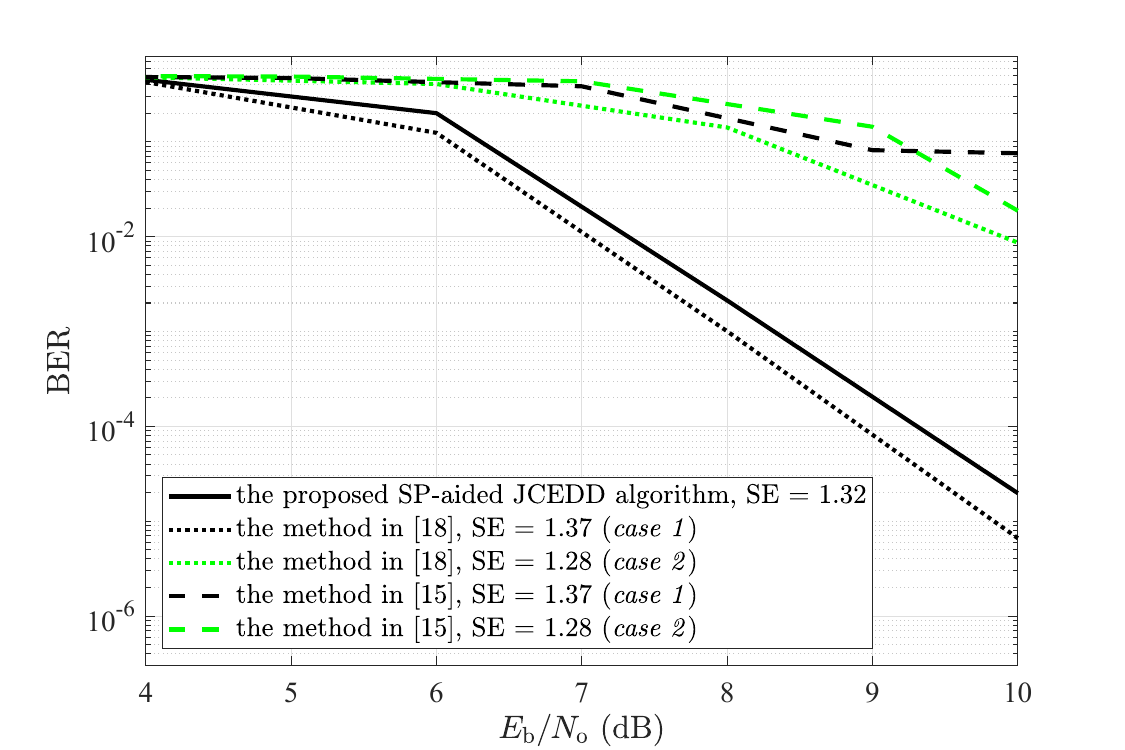} 
\caption{{The BER performance of the proposed SP-aided JCEDD approach for FTN signaling %with an SE of around $1.3$ bits/s/Hz 
over \textit{channel model 2} having $\alpha = 12\%$, with all methods simulated at their third iteration.}}
\label{fig:1224}
\end{figure}

\subsection{BER Results}

{Fig. \ref{fig:121} and Fig. \ref{fig:121it} depict the BER performance of our proposed SP-aided JCEDD method for FTN signaling. The simulation is conducted with a power allocation factor $\alpha = 31\%$ and $\tau = 0.9$ over \textit{channel model 1}. As shown in Fig. \ref{fig:121} and Fig. \ref{fig:121it}, the BER significantly improves after three iterations. More specifically, for $E_\text{b}/N_0$ of 19 dB, the BER approaches $2 \times 10^{-5}$. After four iterations, the BER further decreases, reaching $10^{-5}$ at the same $E_\text{b}/N_0$ level.}

%%%%%%%%%%%%%%%%%%%%%%%%%%%%%

%%%%%%%%%%%%%%%%%%%%%%%%%%%%%%%%%%%%%%
Fig. \ref{fig:122} presents a comparison of the BER performance between our proposed SP-aided SCEDD and JCEDD algorithms with power allocation factor $\alpha = 31\%$ for FTN signaling having $\tau = 0.9$ over \textit{channel model 1} and the MP-aided {approaches employed in \cite{keykhosravi2023pilot} and \cite{ishihara2017iterative}} with all simulations representing the fourth iteration for these methods.
{To maintain fairness, we evaluate two configurations, \textit{case A} and \textit{case B}, for MP-aided transmission over \textit{channel model 1}, as described earlier.} 
As illustrated in Fig. \ref{fig:122}, our proposed SP-aided SCEDD algorithm delivers a significantly better BER compared to the approach presented in \cite{keykhosravi2023pilot} {and \cite{ishihara2017iterative} for both \textit{case A} and \textit{case B}. Furthermore, the BER performance of our SP-aided JCEDD approach surpasses all of} the aforementioned methods by a considerable margin.

%%%%%%%%%%%%%%%%%%%%%%%%%%%%%%%%%%%%%%%%%%
{Fig. \ref{fig:123} offers comparative results of the BER performance of our proposed SP-aided JCEDD approach for FTN signaling with $\tau = 0.9$ and $\tau = 0.8$}, over \textit{channel model 1}. The analysis is conducted under three different power allocation factors, and the number of iterations for the simulations {is set to 4. For a given value of power allocation factor, $\alpha$, the BER experiences lower values for a larger value of $\tau$.}
Fig. \ref{fig:123} reveals that for $\tau = 0.9$, the BER performance is superior when $\alpha = 31\%$, compared to the $\alpha = 21\%$ and $\alpha = 41\%$. This outcome can be attributed to a tradeoff: while a higher power allocation factor results in a lower MSE for channel estimation, it does not always translate to a better BER. Consequently, for a given value of $\tau$, the optimal power allocation factor, in this case $\alpha = 31\%$, should be carefully selected to achieve the best BER performance.
Similarly, Fig. \ref{fig:123} illustrates that for $\tau = 0.8$, the BER performance is better when $\alpha = 41\%$ compared to the $\alpha = 21\%$ and $\alpha = 31\%$. This can be attributed to the increased FTN-induced ISI at $\tau = 0.8$, which leads to less accurate channel estimation for lower power allocation factors, such as $\alpha = 21\%$ and $\alpha = 31\%$. Although lower $\alpha$ values allocate more power to data transmission, the resulting inaccuracy in channel estimation negatively impacts the BER, as highlighted by the MSE results in Fig. \ref{fig:23}.

%%%%%%%%%%%%%%%%%%%%%%%%%%%%%%%%%%%%%
{Fig. \ref{fig:1234} shows the BER performance of our proposed SP-aided JCEDD approach for FTN signaling with $\tau = 0.9$, over \textit{channel model 1}, at three practical $E_\text{b}/N_0$ levels. The number of iterations for the simulations is set to four. As previously discussed, the power allocation factor $\alpha$ must be carefully selected to achieve an effective BER. For a given $\tau$, analyzing BER as a function of $\alpha$ helps determine the optimal value that minimizes BER. 
 Fig. \ref{fig:1234} demonstrates that $\alpha = 31 \%$ yields the lowest BER across all three all three $E_\text{b}/N_0$ levels. Deviating significantly from this value in either direction leads to BER degradation, confirming the existence of a tradeoff in power allocation.
It should be noted that a finer search over $\alpha$ with smaller step sizes could refine the selection. However, since $\alpha$ is a continuous variable, exhaustive search cannot guarantee a globally optimal value. Instead, our chosen $\alpha$ represents a well-balanced tradeoff, validated through simulation.}
%%%%%%%%%%%%%%%%%%%%%%%%%%%%%%%%%%%%%

Fig. \ref{fig:125} illustrates a comparative evaluation of the BER performance for our proposed SP-aided JCEDD approach in FTN signaling with $\tau = 0.8$, $\tau = 0.9$, and $\tau = 1$ (the Nyquist case) over \textit{channel model 1}{, considering two roll-off factor values: $\beta = 0.35$ and $\beta = 0.22$.} The simulation for Fig. \ref{fig:125} is conducted with four iterations, and {the power allocation factor $\alpha$ is carefully selected for each value of $\tau$ to achieve a reasonable BER performance, as indicated in the Fig. \ref{fig:125}}.
As expected, {for a fixed value of $\beta$,} the BER improves as $\tau$ increases. For {$\beta = 0.35$ and} $\tau = 0.9$, Fig. \ref{fig:125} shows a degradation about 1 dB for the $E_\text{b}/N_0$ above 11 dB compared to the $\tau = 1$ (Nyquist case). However, with the $E_\text{b}/N_0$ increasing to 19 dB, the BER of $\tau = 0.9$, reaching $10^{-5}$ and very close to the case of $\tau = 1$ (Nyquist case). The SE for the traditional Nyquist case is equal to 1.11 bits/s/Hz. This {is} an $11\%$ enhancement in SE when employing our proposed SP-aided JCEDD approach. 
{As detailed in the discussion of Fig. \ref{fig:25}, we approximate the delay spread of \textit{channel model 1} to 2.917 ms under Nyquist signaling to ensure a fair comparison with FTN signaling. Under these conditions, variations in $\beta$ do not affect MSE and BER performance in Nyquist signaling.}
{Fig. \ref{fig:125} reveals that reducing $\beta$ in FTN signaling degrades BER, as a slower RRC pulse decay gives rise to ISI at sampling points, worsening performance even for a fixed $\tau$. 
This effect intensifies as $\tau$ decreases; for instance, reducing $\beta$ from 0.35 to o.22 causes minor BER degradation at $\tau = 0.9$ but leads to an approximately 4 dB loss at $\tau = 0.8$ for $E_\text{b}/N_0$ above 18 dB. This occurs because a lower $\tau$ increases pulse overlap, making $\beta$ a more critical factor in determining ISI levels, as discussed earlier.}

%%%%%%%%%%%%%%%%%%%%%%%%%%%%%%%%%%%%%%
Fig. \ref{fig:1224} compares the BER performance of our proposed SP-aided JCEDD algorithm, with a power allocation factor $\alpha = 12\%$, over \textit{channel model 2}, against the MP-aided {methods from \cite{keykhosravi2023pilot} and \cite{ishihara2017iterative}, with all methods} simulated at their third iteration.
{
To ensure fairness, we consider two configurations, \textit{case 1} and \textit{case 2} for MP-aided transmission over \textit{channel model 2}, as previously described.
} {As shown in Fig. \ref{fig:1224}, increasing $N_\text{s}$ (\textit{case 2}) in \cite{keykhosravi2023pilot} degrades BER performance more severely than reducing $\tau$ (\textit{case 1}), despite the latter yielding a slightly higher SE.
Fig. \ref{fig:1224} shows that the SP-aided JCEDD algorithm achieves a BER improvement of more than 2.5 dB over \cite{keykhosravi2023pilot} \textit{case 2} for $E_\text{b}/N_0$ above 7 dB. However, its BER remains comparable to the MP-based approach in \cite{keykhosravi2023pilot} \textit{case 1}, with only a minor degradation of less than 0.5 dB for $E_\text{b}/N_0$ above 8 dB.} This slight degradation is attributed to the low {fading rate} of the \textit{channel model 2}, indicating very slow channel variations that can be tracked more effectively with interpolation in an MP-aided channel estimation method introduced in \cite{keykhosravi2023pilot}. {However, as shown in Fig. \ref{fig:1224}, BER performance of our proposed SP-aided JCEDD algorithm outperforms the method in \cite{ishihara2017iterative} by over 3 dB for $E_\text{b}/N_0$ above 7 dB for both \textit{case 1} and \textit{case 2}.}

The BER results {in Fig. \ref{fig:1224} and Fig. \ref{fig:122}} reveal that for {low fading rates on the order of $10^{-4}$}, the BER of our proposed SP-aided JCEDD algorithm performs similarly to the MP-aided FTN system in \cite{keykhosravi2023pilot}, albeit with slight degradation. However, {for higher fading rates on the order of $10^{-3}$}, our SP-aided method significantly outperforms the MP-aided system, as the MP-aided FTN system in \cite{keykhosravi2023pilot} fails to keep up with the higher fading rates.

\section{Conclusion}

This paper proposed a novel SP-aided approach of doubly-selective channel estimation for FTN signaling. The proposed method is designed to improve the SE. In order to avoid the complexities associated with channel tracking, a BEM is applied, capturing the time-varying properties of doubly-selective channels. More specifically, the time-varying tap weights are represented through a linear combination of basis functions with fixed, time-independent coefficients. We employed a combination of the DTFT and the LSSE method to accurately estimate channel coefficients at each received symbol in the time domain.
Furthermore, we determined the optimal FTN signaling pilot sequence that minimizes the MSE of the doubly-selective channel estimation.
Building on our SP-aided doubly-selective estimation method for FTN signaling, we introduced an SP-aided SCEDD method utilizing a turbo equalizer for detection.
Additionally, to further enhance BER and MSE, we proposed an SP-aided JCEDD method for FTN signaling. 
In this approach, each turbo equalizer iteration incorporates updated channel estimation, which is derived from the data bit estimates within the same iteration and fed back to the equalizer, refining the equalization and decoding accuracy over successive iterations.
Our simulation results indicated that, at the same SE, the SP-aided SCEDD method proposed for FTN signaling achieves significantly better performance in terms of MSE and BER compared to the MP-aided method outlined in \cite{keykhosravi2023pilot} {and \cite{ishihara2017iterative}} for {for higher fading rates on the order of $10^{-3}$}. Additionally, under equivalent SE conditions, our proposed SP-aided JCEDD method for FTN signaling outperforms these methods in both MSE and BER. More specifically, the {methods introduced in the \cite{keykhosravi2023pilot} and \cite{ishihara2017iterative} fail} to track the high fading rate for the practical high-latitude channel introduced in ITU-R F.1487 recommendations, having a Doppler frequency of 10 Hz (fading rate of 0.004). While our simulation results show that our proposed SP-aided JCEDD method for FTN signaling reliably performs under the aforementioned HF channel. At very low {fading rates}, specifically for the practical low-latitude channel defined in the ITU-R F.1487 recommendations, with a Doppler frequency of 1 Hz (fading rate of 0.0004), our SP-aided JCEDD algorithm delivers over 2 dB { and  6 dB improvement in MSE compared to the methods in \cite{keykhosravi2023pilot} and \cite{ishihara2017iterative}, respectively}. However, {while BER performance of our SP-aided JCEDD algorithm shows over 3 dB improvements compared to the MP-aided method in \cite{ishihara2017iterative}, it remains comparable to the MP-aided FTN system in \cite{keykhosravi2023pilot}}, showing only less than 0.5 dB degradation.

\bibliographystyle{IEEEtran}
\bibliography{references}

\end{document}